\documentclass[aps,prl,reprint,superscriptaddress,floatfix]{revtex4-2}
\usepackage{amsmath,amssymb,bm,graphicx}
\usepackage{xcolor}
\usepackage[colorlinks=true,linkcolor=blue!50!black,citecolor=blue!50!black,urlcolor=blue!50!black]{hyperref}
\graphicspath{{figures/}}
\newcommand{\Tr}{\operatorname{Tr}}
\newcommand{\sgn}{\operatorname{sgn}}
\newcommand{\id}{\openone}
\newcommand{\bsig}{\bm{\sigma}}

\newcommand{\Ssec}[1]{SM~\ref{S-#1}}

\let\origaddcontentsline\addcontentsline
\newcommand{\Meq}[1]{Eq.~(\ref{eq:#1})}
\newcommand{\Mfig}[1]{Fig.~\ref{fig:#1}}
\newcommand{\thm}[1]{\par\medskip\noindent\textbf{#1}\ }
\begin{document}

\title{Imaging the topological sign of a Dirac mass with spin-polarized tunneling}

\author{Payman Kazemikhah}
\email{payman@princeton.edu}
\affiliation{Laboratory for Topological Quantum Matter and Advanced Spectroscopy, Department of Physics, Princeton University, Princeton, NJ, USA.}
\affiliation{Department of Chemistry, Princeton University, Princeton, NJ, USA.}
\author{Ruolun Zhang}
\affiliation{Laboratory for Topological Quantum Matter and Advanced Spectroscopy, Department of Physics, Princeton University, Princeton, NJ, USA.}
\author{Byunghoon Kim}
\affiliation{Laboratory for Topological Quantum Matter and Advanced Spectroscopy, Department of Physics, Princeton University, Princeton, NJ, USA.}
\affiliation{Department of Chemistry, Princeton University, Princeton, NJ, USA.}
\author{Xiangyu Luo}
\affiliation{Laboratory for Topological Quantum Matter and Advanced Spectroscopy, Department of Physics, Princeton University, Princeton, NJ, USA.}
\author{Zachary Ducorsky}
\affiliation{Laboratory for Topological Quantum Matter and Advanced Spectroscopy, Department of Physics, Princeton University, Princeton, NJ, USA.}
\author{Yu-Peng Zhu}
\affiliation{Laboratory for Topological Quantum Matter and Advanced Spectroscopy, Department of Physics, Princeton University, Princeton, NJ, USA.}
\author{Md. Shafayat Hossain}
\affiliation{Quantum Materials Intelligence \& Nano Devices (QMIND) Lab, Department of Materials Science and Engineering, University of California Los Angeles (UCLA), CA, USA.}
\author{M. Zahid Hasan}
\email{mzhasan@princeton.edu}
\affiliation{Laboratory for Topological Quantum Matter and Advanced Spectroscopy, Department of Physics, Princeton University, Princeton, NJ, USA.}
\date{\today}

\begin{abstract}
A Dirac mass is characterized not only by its magnitude but also by its sign, which influences the Berry curvature, Hall response, and the propagation direction of chiral boundary modes. However, local spectroscopy of magnetic topological surfaces has so far accessed only the gap magnitude $|M|$. Here we show how the topological sign of $M$ can be determined locally. We first establish the sign information that spin-resolved tunneling can access and the reference required to obtain an absolute Hall sign. Although spin--orbital entanglement generally obscures the relation between measured spin and Dirac pseudospin, crystalline rotation symmetry alleviates this ambiguity at the band edge. On a $C_3$ axis, an axial tip couples to a fully spin-polarized state whose tunneling asymmetry directly reveals the sign of the occupied-band Berry curvature, independent of orbital composition. We further show that the zeroth Landau level provides an independent mass-sign measurement with an unpolarized tip, while opposite domains and their chiral domain-wall modes provide self-consistency checks. These results pave the way for imaging the locations where the topological character changes within a material, rather than merely identifying where a spectral gap appears.
\end{abstract}

\maketitle
\renewcommand{\addcontentsline}[3]{}

Magnetic order can gap the surface Dirac cone of a three-dimensional topological insulator~\cite{HasanKane2010,QiZhang2011}. The resulting massive Dirac state carries a half-quantized surface Hall response~\cite{Qi2008,Nomura2011} and provides the elementary building block of the quantum anomalous Hall effect~\cite{Chang2013,Deng2020,Tokura2019}. Crucially, the gap magnitude does not fully specify this state. The topological sign of the exchange mass $M$ fixes the sign of the occupied-band Berry curvature and surface Hall conductivity, and reversing $M$ reverses the chirality of the mode bound to a magnetic domain wall~\cite{Jackiw1976,Yasuda2017}. Spatial variations of $M$ can therefore partition a nominally uniform sample into regions with opposite local Hall response, connected by networks of chiral channels that can influence macroscopic Hall quantization~\cite{Lee2015}. Determining $M(\mathbf r)$ with its sign would directly reveal this real-space topological structure.

Existing local spectroscopy does not provide that information. Scanning tunneling microscopy (STM) can resolve the magnetic Dirac gap $|M(\mathbf r)|$ with atomic-scale spatial resolution~\cite{Yin2021,Lee2015}, but the spectrum is identical for $+M$ and $-M$. Spin-polarized STM adds sensitivity to the local spin texture~\cite{Wiesendanger2009}, yet the physical spin of a topological surface state is not, in general, the Dirac pseudospin. Spin--orbital entanglement reduces the measured polarization and can even reverse its sign relative to the pseudospin texture~\cite{Yazyev2010,Zhang2013,Jozwiak2013}. Thus, a measured spin contrast cannot by itself be identified with the sign of the Dirac mass. Conversely, transport and optical measurements can determine Hall or magneto-optical signs, but generally average over many microscopic domains. Spin-resolved photoemission resolves the out-of-plane spin texture at the magnetic gap~\cite{Xu2012} and probes symmetry through matrix elements, but it averages over the beam spot, accesses only occupied states, and generally cannot be performed in an applied magnetic field. The missing capability is a local measurement that distinguishes two spectroscopically identical gaps carrying opposite Berry curvature.

Here we establish how tunneling spectroscopy can make that distinction. We first determine, from symmetry, the complete set of mass-sign combinations accessible to spin-resolved tunneling and the external reference required to assign an absolute Hall sign. We then show that crystalline rotation symmetry, whose eigenvalues are known to fix Chern numbers up to multiples of the rotation order~\cite{Fang2012}, removes the principal ambiguity from spin--orbital entanglement: on a $C_3$ axis, the band-edge state seen by an axial tip is fully spin-polarized, so the sign of the tunneling asymmetry directly gives the sign of the occupied-band Berry curvature, independent of orbital composition. An independent route follows from the zeroth Landau level, whose position relative to the gap determines $\sgn(MB)$ even with an unpolarized tip, or from the field-odd shift of the band edges when Landau levels are not resolved. Opposite magnetic domains and their domain-wall modes provide internal consistency tests, while explicit noise and tip-perturbation bounds define the experimental requirements. Together, these results provide a route to imaging not only where a Dirac surface is gapped, but also where its local topological character changes in real space. Derivations and numerical tests are given in the Supplemental Material (SM)~\cite{SM}.

\begin{figure*}[t]
\includegraphics[width=0.96\textwidth]{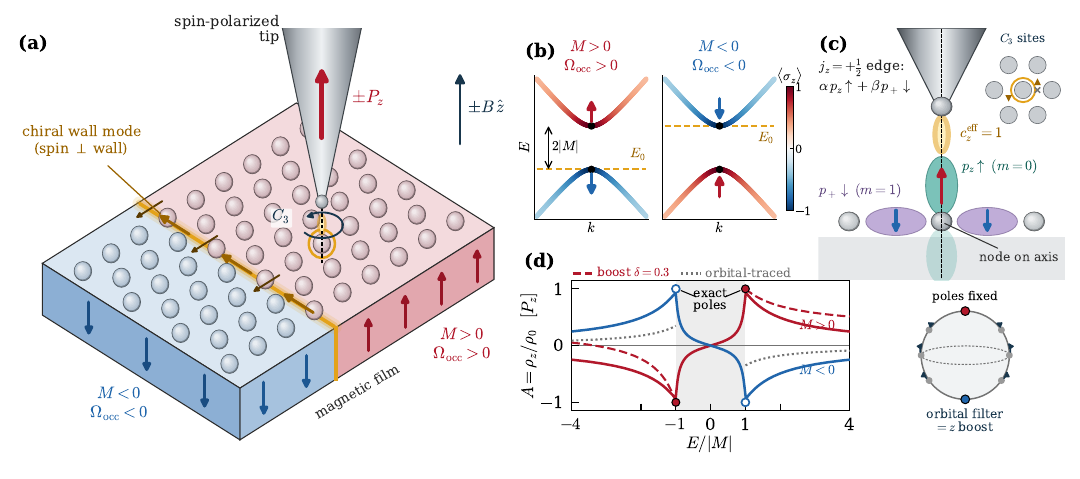}
\caption{\textbf{Local readout of the Dirac-mass sign.}
(a)~A spin-polarized tip with reversible polarization $\pm P_z$ probes a $C_3$ axis of a magnetic topological-insulator surface. Domains with opposite exchange masses $M$ have opposite Berry curvatures, and their boundary hosts a chiral mode with in-plane spin perpendicular to the wall.
(b)~Massive Dirac spectra for the two domains, colored by $\langle\sigma_z\rangle$. The band-edge spin reverses with $M$; the dashed line marks the zeroth Landau level, $E_0=-M$ for $eB>0$.
(c)~Rotation-axis selection rule. For the band-edge state $\alpha p_z{\uparrow}+\beta p_+{\downarrow}$, the $m=1$ component vanishes on the $C_3$ axis, so an axial tip couples to a fully polarized state, $c_z^{\rm eff}=1$, independent of the orbital weights $|\alpha|^2$ and $|\beta|^2$. Inset: symmetry-equivalent $C_3$ axes at atop, hcp, and fcc sites.
(d)~Spin asymmetry $A(E)$ for opposite mass domains. The band-edge values, $\pm P_z\,\sgn\Omega_{\rm occ}$, are fixed by the Berry-curvature sign. Orbital filtering away from the edge acts as the pseudospin transformation of Eq.~\eqref{eq:boost} (dashed, $\delta=0.3$; Bloch-circle representation at right) and leaves the band-edge values unchanged, whereas the orbital-traced spin polarization (dotted, $c_z=-0.35$) can have the opposite sign.}
\label{fig:setup}
\end{figure*}

The surface Hamiltonian is
\begin{equation}
H=v(k_x\sigma_y-k_y\sigma_x)+\bigl[M-\beta k^2+\lambda\,\mathrm{Re}\,k_+^3\bigr]\sigma_z ,
\label{eq:H}
\end{equation}
with warping $\lambda$~\cite{Fu2009}, curvature $\beta$, and $E=\sqrt{v^2k^2+M^2}$ for $\lambda=\beta=0$. The Pauli matrices act on the Kramers doublet at $\bar\Gamma$, the pseudospin. They act on the spin only for trivial orbital content. A tip with polarization $\mathbf P$ has effect $W=g(\id+\mathbf P\cdot\bsig)$ and measures $\rho_W=g(\rho_0+\mathbf P\cdot\bm\rho)$, with $\rho_a=-\pi^{-1}\mathrm{Im}\Tr[\sigma_aG^R(\mathbf r,\mathbf r;E)]$ [Fig.~\ref{fig:setup}(a)]. We define the mass sign as the sign of the occupied-band Berry curvature, $\Omega_{\rm occ}=v^2M/2E^3$ for either sign of $v$ (\Ssec{sec:conventions}). This definition is basis independent, and it equals the sign of the half-quantized Hall response [Fig.~\ref{fig:setup}(b)].

\begin{table}[b]
\caption{Signs determined by spin-resolved tunneling with tip polarizations $P_\parallel$ and $P_z$, defect moments $u$, and a perpendicular field $B$. ``Yes'' means that the absolute Hall sign follows once the reference is calibrated in the laboratory frame.}
\label{tab:signs}
\begin{ruledtabular}
\begin{tabular}{p{0.34\columnwidth}p{0.31\columnwidth}p{0.19\columnwidth}}
\raggedright Reference & \raggedright Determined & \raggedright $\sgn M$?\tabularnewline
\hline
\raggedright unpolarized, scalar, $B=0$ & \raggedright nothing & \raggedright no\tabularnewline
\raggedright in-plane tip or moment & \raggedright $\sgn(vMP_\parallel)$ & \raggedright with $\sgn v$\tabularnewline
\raggedright $z$ tip ($z$ moment) & \raggedright $\sgn(MP_z)$ [$\sgn(Mu_z)$] & \raggedright yes\tabularnewline
\raggedright field $B\hat{\mathbf z}$ & \raggedright $\sgn(MB)$ & \raggedright yes, \mbox{unpolarized}\tabularnewline
\raggedright in-plane and $z$ tips & \raggedright also $\sgn(vP_\parallel P_z)$ & \raggedright yes, and $\sgn v$
\end{tabular}
\end{ruledtabular}
\end{table}

Three sign reversals leave the problem invariant as a whole. They are $\Theta_\perp=\sigma_xK$, time reversal $\Theta_T=i\sigma_yK$, and $R_z=i\sigma_z$. For antiunitary $\Theta$, $G^R_{\Theta H\Theta^{-1}}=\Theta G^A_H\Theta^{-1}$. Every map of $\Theta H\Theta^{-1}$ taken with tip $W$ therefore equals the map of $H$ taken with tip $\Theta^{-1}W\Theta$. This holds for any defect, for the exact $T$ matrix, and for any broadening (\Ssec{sec:sign}).

The three operations, together with the identity, form a group. They reverse the signs of $(v,M,B,P_z)$, $(M,B,P_\parallel,P_z)$, and $(v,P_\parallel)$, respectively. The warping $\lambda$ is invariant, and $\beta$ is reversed together with $M$. Only three independent sign combinations survive all three operations, $\sgn(MP_z)$, $\sgn(vMP_\parallel)$, and $\sgn(MB)$. Tunneling data can therefore determine at most these three signs, and each is fixed by its own reference (Table~\ref{tab:signs}). In-plane channels have the backscattering coherence factor $\Tr[\sigma_xP(\mathbf k)P(-\mathbf k)]=in_yn_z\propto vM$. They determine $\sgn(vM)$, not the mass sign. An adatom with a known out-of-plane moment would itself be a $z$ reference. Without a time-reversal-odd reference, static maps encode no chirality, however chiral their patterns look.

Time-reversal symmetry forces the spin signal of a nonmagnetic surface to vanish. If $\Theta_TH\Theta_T^{-1}=H$, with arbitrary disorder, orbitals, and spin--orbit coupling, then
\begin{equation}
\Tr_{\rm orb}\bigl[\sigma_a\,G^R(\mathbf R,\mathbf R;E)\bigr]=0
\label{eq:kramers}
\end{equation}
identically, for the real and imaginary parts. A nonmagnetic reference therefore shows an exactly vanishing spin asymmetry. Time reversal also reverses $B$, so the $\pm B$-averaged spin signal is exactly odd in the magnetization, to all orders (\Ssec{sec:kramers}).

The spin-resolved density of states is a contour average weighted by the density of states, $\rho_a/\rho_0=\langle\pm n_a\rangle_E$, where $\mathbf n$ is the upper-band pseudospin. For the isotropic cone, $A\equiv\rho_z/\rho_0=M/E$ at signed energy $E$ with $|E|>|M|$ [Fig.~\ref{fig:setup}(d)]. With Lorentzian broadening $\eta$ the numerator is free of cutoffs,
\begin{equation}
\rho_z(E)=\frac{M}{2\pi^2v^2}\left[\arctan\frac{E-|M|}{\eta}+\arctan\frac{E+|M|}{\eta}\right],
\label{eq:closed}
\end{equation}
while $\rho_0$ gains only an $\eta\ln\Lambda$ correction. Hence $A(\pm|M|)\to\pm\sgn M$ as $\eta\to0$ (\Ssec{sec:local}). Warping and curvature create no signal on a gapless surface. The ``band-inversion'' sign change of lattice-regularized models is an artefact of their time-reversal-odd curvature term.

Spin--orbit coupling creates the surface Dirac cone and entangles spin with orbital motion, so the pseudospin is not the electron spin. On a real surface the Dirac doublet lives in spin $\otimes$ orbital space. The orbital-traced spin polarization is reduced~\cite{Yazyev2010} and can change sign~\cite{Zhang2013,Jozwiak2013}. The relation $A=P_zc_zM/E$ would then seem to require $\sgn c_z$ from first principles. At the band edge, however, the tip does not see $c_z$.

Rotation symmetry removes this obstacle on the rotation axes. Consider a surface with a $C_n$ axis ($n\ge2$) and a magnetization along it, and a band-edge state $\psi$ at $\bar\Gamma$ with rotation eigenvalue $e^{-2\pi ij_z/n}$, $j_z=\pm\frac12$. On the axis the rotation acts only on the spin. The spin components therefore obey $e^{-2\pi is_z/n}\psi_s=e^{-2\pi ij_z/n}\psi_s$, so $\psi_s=0$ unless $s_z=j_z$. On every $C_n$ axis the state is thus fully spin polarized, with $\psi_\downarrow=0$ for $j_z=+\frac12$ and $\psi_\uparrow=0$ for $j_z=-\frac12$. All $C_n$ axes share this eigenvalue at $\bar\Gamma$, because they differ by lattice translations. The same holds for $D\psi$, with $D$ any derivative that commutes with rotations about the axis, which covers $s$, $p_z$, and $d_{z^2}$ tips~\cite{Chen1990}. The argument uses only rotation symmetry, so it holds for any strength of spin--orbit coupling.

An axial tip on atop or hollow sites thus measures $c_z^{\rm eff}=1$ exactly, for any orbital composition [Fig.~\ref{fig:setup}(c)]. The same argument, applied to the laterally averaged spinor, shows that the $\mathbf G=0$ vacuum component of the band-edge state is spin pure. The $\mathbf G\ne0$ components decay faster, so $1-|c_z^{\rm eff}|\propto e^{-2\Delta\kappa z}$ everywhere, with $\Delta\kappa=\sqrt{\kappa^2+G^2}-\kappa$. For Bi$_2$Se$_3$ ($G=1.75$~\AA$^{-1}$, $\kappa\approx1.1$~\AA$^{-1}$) this is about $10^{-4}$ at $z=5$~\AA\ (\Ssec{sec:selection}).

For $n\ge3$, rotation symmetry makes the linear term between the $j_z=\pm\frac12$ states proportional to $k_-$. Its winding is $+1$ for every orbital composition. With $H_{\rm eff}=\tilde M\tau_z+\dots$ in the $j_z$ basis, $\sgn\Omega_{\rm occ}=\sgn\tilde M$, and the upper edge has $j_z=\frac12\sgn\tilde M$. Hence
\begin{equation}
A_{\rm axis}(E\to\pm|\tilde M|)=\pm P_z\,\sgn\Omega_{\rm occ}
\label{eq:axis}
\end{equation}
exactly, for any linearly dispersing surface. To first order $\tilde M=Jc_z$. The orbital-traced spin of the upper edge thus has the sign of the exchange field $J$, and the on-axis signal has the sign of $Jc_z$. Their ratio measures $\sgn c_z$. If the magnetization is known from an independent probe, the on-axis measurement thus fixes the material constant $\sgn(Jc_z)$. The local measurement remains essential where $J$ or $c_z$ varies, for example at antisite defects, terminations, and interfaces. Away from the edge the tip weighs the $j_z$ components with unequal weights $w_\pm$. The measured Bloch vector is then a Lorentz boost of the pseudospin along $z$,
\begin{equation}
A_{\rm axis}(E)=P_z\frac{n_z(E)+\delta}{1+\delta\,n_z(E)},\quad \delta=\frac{w_+-w_-}{w_++w_-}=O(J/\Delta_r),
\label{eq:boost}
\end{equation}
with $\Delta_r$ the distance to remote bands. The poles $n_z=\pm1$ are fixed points. Filtering can therefore shift the signal away from the edge, but never at it [Fig.~\ref{fig:setup}(d)].

\begin{figure*}[t]
\includegraphics[width=0.9\textwidth]{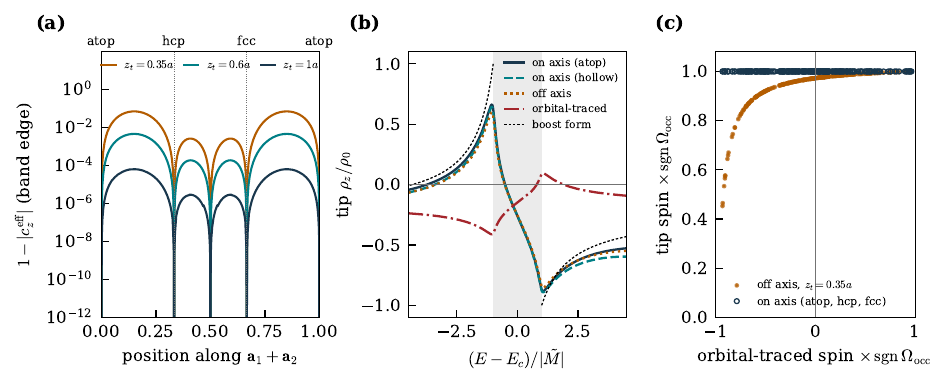}
\caption{\textbf{On-axis spin equals the Hall sign.}
Selection rule in a $C_{3v}$ spin--orbital model ($p$ orbitals, spin--orbit and orbital Rashba coupling, exchange along $z$, \Ssec{sec:selection}). (a) Deviation $1-|c_z^{\rm eff}|$ of the tip polarization of the upper band edge along the cell diagonal for three tip heights. It vanishes on the $C_3$ axes. The zero at the bridge site reflects the model's twofold axis. (b) Tip asymmetry near the gap on and off the axis, and orbital-traced. Dashed, Eq.~\eqref{eq:boost} with $\delta=-0.23$. The orbital-traced polarization has the \emph{opposite} sign at both edges ($+0.16$, $-0.48$). (c) For 250 random parameter sets, the on-axis spin of the upper edge always equals $\sgn\Omega_{\rm occ}$. The orbital-traced spin has either sign.}
\label{fig:selection}
\end{figure*}

Figure~\ref{fig:selection} tests these statements in a $C_{3v}$ tight-binding model of $p$ orbitals with an $s$-wave tip. The orbital-traced polarization of the upper edge is $+0.16$. A tip on any $C_3$ axis sees exactly $-1$, the sign of $\Omega_{\rm occ}$, and $\tilde M/Jc_z=0.9999$ at small $J$. In 9872 random parameter sets the on-axis sign equals the Hall sign in every case, while the orbital-traced sign is opposite in 57\% of them. The deviation from unity grows quadratically with lateral offset and tilt and decays with height at the predicted rate. Tip orbitals with $m\ne0$, such as the $p_x$ and $p_y$ orbitals of CO-functionalized tips, see the minority spin on the axis, but they couple only to the suppressed $\mathbf G\ne0$ components. Their contribution decays faster with height, so a tip-height series exposes it. The Landau route below reads an energy rather than a spin and is immune to it.

\begin{figure*}[t]
\includegraphics[width=0.9\textwidth]{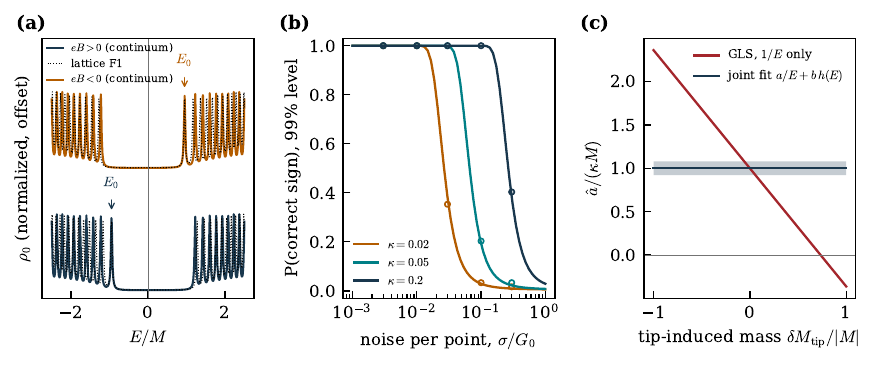}
\caption{\textbf{Field route, decision statistics, and tip-induced mass.}
(a) Landau levels for $eB>0$ (lower) and $eB<0$ (upper) for the continuum with curvature (solid) and lattice F1 (dotted), $M=0.3$, $2\pi\phi=0.031$, with $E_0=-\sgn(eB)(M-\beta|eB|)$. (b) Probability of a correct decision at 99\% confidence, Eq.~\eqref{eq:power} (lines) and Monte Carlo (circles), $M=15$~meV. (c) Bias from a tip-induced mass in fits that use one side of the gap, $1/E$ only versus joint fit. The band marks $\pm1$ standard error at $\sigma=0.01G_0$.}
\label{fig:stats}
\end{figure*}

In a perpendicular field, with $\bm\pi=\mathbf p+e\mathbf A$, the operator $a=(\pi_x-i\pi_y)/\sqrt{2\hbar eB}$ lowers the Landau index for $eB>0$. Equation~\eqref{eq:H} with $\lambda=0$ then has the zero mode $(0,|0\rangle)$ at
\begin{equation}
E_0=-\sgn(eB)\,\bigl(M-\beta|eB|/\hbar\bigr),
\label{eq:E0}
\end{equation}
independent of $v$. All other levels lie at $\pm[M_n^2+2n\hbar v^2|eB|]^{1/2}$ with $\langle\sigma_z\rangle=M_n/E_n$ (\Ssec{sec:landau}). The zeroth level sits at the lower band edge when $\sgn(eBM)>0$ and at the upper edge otherwise [Fig.~\ref{fig:setup}(b)]. By Streda's relation~\cite{Streda1982} this gives the Hall sign with an \emph{unpolarized} tip, referenced to the laboratory field [$\sgn(MB)$ in Table~\ref{tab:signs}]. Lattice calculations confirm Eq.~\eqref{eq:E0} to $3\times10^{-4}$ [Fig.~\ref{fig:stats}(a)]. For $v=4\times10^5$~m/s the level spacing is $14.5$~meV\,$\sqrt{B[\mathrm T]}$, and $\ell_B=25.7$~nm$/\sqrt{B[\mathrm T]}$ sets the resolution. When Landau levels are not resolved, for instance in doped or disordered samples, the same sign appears in the field-odd shift of the band edges. Both edges move with the orbital magnetic moment of the band edge, so the gap midpoint shifts by $\sgn(eBM)\hbar v^2|eB|/2|M|$, about 3.5~meV per tesla for $|M|=15$~meV, while the Zeeman term changes the size of the gap but not its midpoint. The link between this orbital moment and the Berry curvature is exact for an isolated two-band edge. In 3993 random multi-orbital models with gapped band edges, the sign of the shift equals the Hall sign in every case (\Ssec{sec:landau}).

\begin{figure*}[t]
\includegraphics[width=0.9\textwidth]{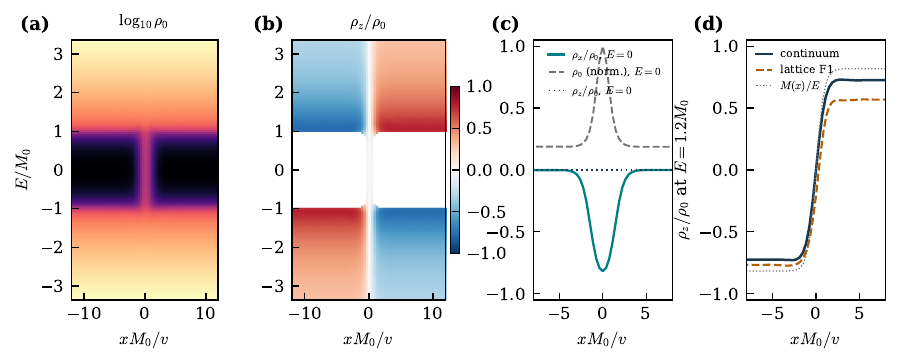}
\caption{\textbf{Domain walls as self-calibrating controls.}
Domain walls in the doubler-free continuum model ($M_0=0.3$, wall width $0.9\,v/M_0$, $\eta=0.1M_0$). (a) $\log_{10}\rho_0$ with the in-gap Jackiw--Rebbi mode. (b) $\rho_z/\rho_0$, exactly antisymmetric about the wall. (c) At $E=0$ the mode has spin perpendicular to the wall ($\rho_x/\rho_0=-0.82$) and is invisible to a $z$ tip. (d) Cuts at $E=1.2M_0$ for the continuum ($\pm0.73$) and lattice F1 ($+0.57$, $-0.77$). The asymmetry of F1 comes from its time-reversal-odd regulator.}
\label{fig:walls}
\end{figure*}

The opposite uniform domains are exact time-reversal partners, so their $\pm B$-averaged spin signals are exactly opposite. For a wall with $M(-x)=-M(x)$, the mirror $M_x=i\sigma_x$ maps $M(x)\to-M(-x)$. This gives $\rho_z(x)=-\rho_z(-x)$ exactly [Fig.~\ref{fig:walls}(b), residual $3\times10^{-15}$]. The Jackiw--Rebbi mode~\cite{Jackiw1976} has spin $\sigma_x=-\sgn(v\,\partial_xM)$ perpendicular to the wall and energy $E=|v|k_y\sgn(\partial_xM)$. It is dark for $z$ tips and bright for in-plane tips [Fig.~\ref{fig:walls}(c)]. Seen from vacuum, the boundary mode of a domain with $\Omega_{\rm occ}>0$ circulates clockwise. The sign flip, the exact antisymmetry, and the in-gap mode together form a self-calibrating control. A time-reversal-odd lattice regulator breaks this antisymmetry, while an anomaly-free slab keeps it (\Ssec{sec:walls}).

For a signal $A_j=\kappa M/E_j$ with noise $\sigma$ at energies $E_j$, the best linear fit gives the amplitude $\hat a=S_2^{-1}\sum_jA_j/E_j$, with $S_2=\sum_jE_j^{-2}$. The sign is then correctly assigned, at the confidence level $1-\alpha$, with probability
\begin{equation}
P_{\rm correct}=\Phi\bigl(|\kappa M|\sqrt{S_2}/\sigma-z_{1-\alpha/2}\bigr)
\label{eq:power}
\end{equation}
[Fig.~\ref{fig:stats}(b)]. For $M=15$~meV, 60 energies in $[1.3,4]|M|$, and $\kappa=0.05$, $P=0.19$ at $\sigma=0.1G_0$ and $P=0.999$ at $\sigma=0.03G_0$. The Zeeman mass of the applied field cancels exactly in the $\pm B$ average. A magnetized tip can also shift the mass under the apex by $\delta M_{\rm tip}\propto P$, through exchange or the Zeeman part of its stray field. By particle--hole symmetry this shift is even in $E$ to first order, while the signal is odd. It therefore cancels when the upper and lower band edges are compared, to $<0.4\%$ in simulations. It biases only fits that use one side of the gap, by $-27\%$ at $\delta M_{\rm tip}=0.2|M|$ for $\kappa=0.05$ [Fig.~\ref{fig:stats}(c)]. The orbital part of a ferromagnetic tip's stray field is odd and does not cancel. At $\kappa=0.2$ it reverses the sign read near the band edge above about 1.5~T, and the same field can switch the sample. Antiferromagnetic tips avoid both effects (\Ssec{sec:tip}).

We propose a straightforward protocol. (i)~Calibrate $\sgn P_z$ on a reference magnet, and check Eq.~\eqref{eq:kramers} on a nonmagnetic topological insulator. (ii)~Record $\pm B$-averaged tip-reversal spectra on atop or hollow sites at both band edges, preferably with an antiferromagnetic tip. Apply Eq.~\eqref{eq:axis} in windows symmetric about the local gap center, and extract the sign with a joint fit of Eq.~\eqref{eq:boost} or a tip-height series. (iii)~Cross-check with the zeroth Landau level, Eq.~\eqref{eq:E0}, using $\rho_0(+B)-\rho_0(-B)$ within the local half gap, or, where Landau levels are not resolved, with the field-odd shift of the gap midpoint. (iv)~Image domain walls, where the antisymmetry, the in-gap mode, and the in-plane contrast must agree. Compare gap maps from a nonmagnetic tip before and after to exclude switching. Large-scale lattice simulations confirm these statements. Under 30\% mass disorder, the sign read at the band edges and from the field-odd zeroth level is correct at $\ge99\%$ of sites. The band-edge reading also images patches of reversed sign beyond one Dirac length from their boundaries (\Ssec{sec:campaigns}).

Immediate experimental tests are feasible in established magnetic topological materials. MnBi$_2$Te$_4$~\cite{Otrokov2019,Deng2020} and Cr- or V-doped (Bi,Sb)$_2$Te$_3$~\cite{Chang2013,Chang2015} provide natural platforms in which the Dirac gap, magnetic domains, and anomalous Hall response are already experimentally accessible. Gapped Dirac states in kagome magnets such as TbMn$_6$Sn$_6$~\cite{Yin2020} and Fe$_3$Sn$_2$~\cite{Ye2018} suggest a broader class of systems in which analogous symmetry-based readout may be possible. More generally, the same spin-resolved tunneling observables contain information about the Berry curvature and quantum metric. The present framework therefore points beyond identifying the topological sign of a Dirac mass: it opens a route toward resolving quantum geometry and local topological character in real space, with atomic-scale access to where these quantities change across domains, defects, and boundaries.

\begin{acknowledgments}
Advanced ARPES, STM, transport, and theoretical work at Princeton University were supported by the US DOE under the Basic Energy Sciences program (Grant No.~DOE/BES DE-FG-02-05ER46200; M.Z.H.), National Quantum Information Science Research Centers, Quantum Science Center (at ORNL) supported by US DOE and the Gordon and Betty Moore Foundation (No.~GBMF9461; M.Z.H.). The numerical code and data are available from the authors.
\end{acknowledgments}

\bibliography{references}

%apsrev4-2.bst 2019-01-14 (MD) hand-edited version of apsrev4-1.bst
%Control: key (0)
%Control: author (8) initials jnrlst
%Control: editor formatted (1) identically to author
%Control: production of article title (0) allowed
%Control: page (0) single
%Control: year (1) truncated
%Control: production of eprint (0) enabled
\begin{thebibliography}{27}%
\makeatletter
\providecommand \@ifxundefined [1]{%
 \@ifx{#1\undefined}
}%
\providecommand \@ifnum [1]{%
 \ifnum #1\expandafter \@firstoftwo
 \else \expandafter \@secondoftwo
 \fi
}%
\providecommand \@ifx [1]{%
 \ifx #1\expandafter \@firstoftwo
 \else \expandafter \@secondoftwo
 \fi
}%
\providecommand \natexlab [1]{#1}%
\providecommand \enquote  [1]{``#1''}%
\providecommand \bibnamefont  [1]{#1}%
\providecommand \bibfnamefont [1]{#1}%
\providecommand \citenamefont [1]{#1}%
\providecommand \href@noop [0]{\@secondoftwo}%
\providecommand \href [0]{\begingroup \@sanitize@url \@href}%
\providecommand \@href[1]{\@@startlink{#1}\@@href}%
\providecommand \@@href[1]{\endgroup#1\@@endlink}%
\providecommand \@sanitize@url [0]{\catcode `\\12\catcode `\$12\catcode
  `\&12\catcode `\#12\catcode `\^12\catcode `\_12\catcode `\%12\relax}%
\providecommand \@@startlink[1]{}%
\providecommand \@@endlink[0]{}%
\providecommand \url  [0]{\begingroup\@sanitize@url \@url }%
\providecommand \@url [1]{\endgroup\@href {#1}{\urlprefix }}%
\providecommand \urlprefix  [0]{URL }%
\providecommand \Eprint [0]{\href }%
\providecommand \doibase [0]{https://doi.org/}%
\providecommand \selectlanguage [0]{\@gobble}%
\providecommand \bibinfo  [0]{\@secondoftwo}%
\providecommand \bibfield  [0]{\@secondoftwo}%
\providecommand \translation [1]{[#1]}%
\providecommand \BibitemOpen [0]{}%
\providecommand \bibitemStop [0]{}%
\providecommand \bibitemNoStop [0]{.\EOS\space}%
\providecommand \EOS [0]{\spacefactor3000\relax}%
\providecommand \BibitemShut  [1]{\csname bibitem#1\endcsname}%
\let\auto@bib@innerbib\@empty
%</preamble>
\bibitem [{\citenamefont {Hasan}\ and\ \citenamefont
  {Kane}(2010)}]{HasanKane2010}%
  \BibitemOpen
  \bibfield  {author} {\bibinfo {author} {\bibfnamefont {M.~Z.}\ \bibnamefont
  {Hasan}}\ and\ \bibinfo {author} {\bibfnamefont {C.~L.}\ \bibnamefont
  {Kane}},\ }\bibfield  {title} {\bibinfo {title} {Colloquium: Topological
  insulators},\ }\href {https://doi.org/10.1103/RevModPhys.82.3045} {\bibfield
  {journal} {\bibinfo  {journal} {Rev. Mod. Phys.}\ }\textbf {\bibinfo {volume}
  {82}},\ \bibinfo {pages} {3045} (\bibinfo {year} {2010})}\BibitemShut
  {NoStop}%
\bibitem [{\citenamefont {Qi}\ and\ \citenamefont {Zhang}(2011)}]{QiZhang2011}%
  \BibitemOpen
  \bibfield  {author} {\bibinfo {author} {\bibfnamefont {X.-L.}\ \bibnamefont
  {Qi}}\ and\ \bibinfo {author} {\bibfnamefont {S.-C.}\ \bibnamefont {Zhang}},\
  }\bibfield  {title} {\bibinfo {title} {Topological insulators and
  superconductors},\ }\href {https://doi.org/10.1103/RevModPhys.83.1057}
  {\bibfield  {journal} {\bibinfo  {journal} {Rev. Mod. Phys.}\ }\textbf
  {\bibinfo {volume} {83}},\ \bibinfo {pages} {1057} (\bibinfo {year}
  {2011})}\BibitemShut {NoStop}%
\bibitem [{\citenamefont {Qi}\ \emph {et~al.}(2008)\citenamefont {Qi},
  \citenamefont {Hughes},\ and\ \citenamefont {Zhang}}]{Qi2008}%
  \BibitemOpen
  \bibfield  {author} {\bibinfo {author} {\bibfnamefont {X.-L.}\ \bibnamefont
  {Qi}}, \bibinfo {author} {\bibfnamefont {T.~L.}\ \bibnamefont {Hughes}},\
  and\ \bibinfo {author} {\bibfnamefont {S.-C.}\ \bibnamefont {Zhang}},\
  }\bibfield  {title} {\bibinfo {title} {Topological field theory of
  time-reversal invariant insulators},\ }\href
  {https://doi.org/10.1103/PhysRevB.78.195424} {\bibfield  {journal} {\bibinfo
  {journal} {Phys. Rev. B}\ }\textbf {\bibinfo {volume} {78}},\ \bibinfo
  {pages} {195424} (\bibinfo {year} {2008})}\BibitemShut {NoStop}%
\bibitem [{\citenamefont {Nomura}\ and\ \citenamefont
  {Nagaosa}(2011)}]{Nomura2011}%
  \BibitemOpen
  \bibfield  {author} {\bibinfo {author} {\bibfnamefont {K.}~\bibnamefont
  {Nomura}}\ and\ \bibinfo {author} {\bibfnamefont {N.}~\bibnamefont
  {Nagaosa}},\ }\bibfield  {title} {\bibinfo {title} {Surface-quantized
  anomalous {H}all current and the magnetoelectric effect in magnetically
  disordered topological insulators},\ }\href
  {https://doi.org/10.1103/PhysRevLett.106.166802} {\bibfield  {journal}
  {\bibinfo  {journal} {Phys. Rev. Lett.}\ }\textbf {\bibinfo {volume} {106}},\
  \bibinfo {pages} {166802} (\bibinfo {year} {2011})}\BibitemShut {NoStop}%
\bibitem [{\citenamefont {Chang}\ \emph {et~al.}(2013)\citenamefont {Chang}
  \emph {et~al.}}]{Chang2013}%
  \BibitemOpen
  \bibfield  {author} {\bibinfo {author} {\bibfnamefont {C.-Z.}\ \bibnamefont
  {Chang}} \emph {et~al.},\ }\bibfield  {title} {\bibinfo {title} {Experimental
  observation of the quantum anomalous {H}all effect in a magnetic topological
  insulator},\ }\href {https://doi.org/10.1126/science.1234414} {\bibfield
  {journal} {\bibinfo  {journal} {Science}\ }\textbf {\bibinfo {volume}
  {340}},\ \bibinfo {pages} {167} (\bibinfo {year} {2013})}\BibitemShut
  {NoStop}%
\bibitem [{\citenamefont {Deng}\ \emph {et~al.}(2020)\citenamefont {Deng} \emph
  {et~al.}}]{Deng2020}%
  \BibitemOpen
  \bibfield  {author} {\bibinfo {author} {\bibfnamefont {Y.}~\bibnamefont
  {Deng}} \emph {et~al.},\ }\bibfield  {title} {\bibinfo {title} {Quantum
  anomalous {H}all effect in intrinsic magnetic topological insulator
  {MnBi$_2$Te$_4$}},\ }\href {https://doi.org/10.1126/science.aax8156}
  {\bibfield  {journal} {\bibinfo  {journal} {Science}\ }\textbf {\bibinfo
  {volume} {367}},\ \bibinfo {pages} {895} (\bibinfo {year}
  {2020})}\BibitemShut {NoStop}%
\bibitem [{\citenamefont {Tokura}\ \emph {et~al.}(2019)\citenamefont {Tokura},
  \citenamefont {Yasuda},\ and\ \citenamefont {Tsukazaki}}]{Tokura2019}%
  \BibitemOpen
  \bibfield  {author} {\bibinfo {author} {\bibfnamefont {Y.}~\bibnamefont
  {Tokura}}, \bibinfo {author} {\bibfnamefont {K.}~\bibnamefont {Yasuda}},\
  and\ \bibinfo {author} {\bibfnamefont {A.}~\bibnamefont {Tsukazaki}},\
  }\bibfield  {title} {\bibinfo {title} {Magnetic topological insulators},\
  }\href {https://doi.org/10.1038/s42254-018-0011-5} {\bibfield  {journal}
  {\bibinfo  {journal} {Nat. Rev. Phys.}\ }\textbf {\bibinfo {volume} {1}},\
  \bibinfo {pages} {126} (\bibinfo {year} {2019})}\BibitemShut {NoStop}%
\bibitem [{\citenamefont {Jackiw}\ and\ \citenamefont
  {Rebbi}(1976)}]{Jackiw1976}%
  \BibitemOpen
  \bibfield  {author} {\bibinfo {author} {\bibfnamefont {R.}~\bibnamefont
  {Jackiw}}\ and\ \bibinfo {author} {\bibfnamefont {C.}~\bibnamefont {Rebbi}},\
  }\bibfield  {title} {\bibinfo {title} {Solitons with fermion number 1/2},\
  }\href {https://doi.org/10.1103/PhysRevD.13.3398} {\bibfield  {journal}
  {\bibinfo  {journal} {Phys. Rev. D}\ }\textbf {\bibinfo {volume} {13}},\
  \bibinfo {pages} {3398} (\bibinfo {year} {1976})}\BibitemShut {NoStop}%
\bibitem [{\citenamefont {Yasuda}\ \emph {et~al.}(2017)\citenamefont {Yasuda}
  \emph {et~al.}}]{Yasuda2017}%
  \BibitemOpen
  \bibfield  {author} {\bibinfo {author} {\bibfnamefont {K.}~\bibnamefont
  {Yasuda}} \emph {et~al.},\ }\bibfield  {title} {\bibinfo {title} {Quantized
  chiral edge conduction on domain walls of a magnetic topological insulator},\
  }\href {https://doi.org/10.1126/science.aan5991} {\bibfield  {journal}
  {\bibinfo  {journal} {Science}\ }\textbf {\bibinfo {volume} {358}},\ \bibinfo
  {pages} {1311} (\bibinfo {year} {2017})}\BibitemShut {NoStop}%
\bibitem [{\citenamefont {Lee}\ \emph {et~al.}(2015)\citenamefont {Lee} \emph
  {et~al.}}]{Lee2015}%
  \BibitemOpen
  \bibfield  {author} {\bibinfo {author} {\bibfnamefont {I.}~\bibnamefont
  {Lee}} \emph {et~al.},\ }\bibfield  {title} {\bibinfo {title} {Imaging
  {D}irac-mass disorder from magnetic dopant atoms in the ferromagnetic
  topological insulator {Cr$_x$(Bi$_{0.1}$Sb$_{0.9}$)$_{2-x}$Te$_3$}},\ }\href
  {https://doi.org/10.1073/pnas.1424322112} {\bibfield  {journal} {\bibinfo
  {journal} {Proc. Natl. Acad. Sci. USA}\ }\textbf {\bibinfo {volume} {112}},\
  \bibinfo {pages} {1316} (\bibinfo {year} {2015})}\BibitemShut {NoStop}%
\bibitem [{\citenamefont {Yin}\ \emph {et~al.}(2021)\citenamefont {Yin},
  \citenamefont {Pan},\ and\ \citenamefont {Hasan}}]{Yin2021}%
  \BibitemOpen
  \bibfield  {author} {\bibinfo {author} {\bibfnamefont {J.-X.}\ \bibnamefont
  {Yin}}, \bibinfo {author} {\bibfnamefont {S.~H.}\ \bibnamefont {Pan}},\ and\
  \bibinfo {author} {\bibfnamefont {M.~Z.}\ \bibnamefont {Hasan}},\ }\bibfield
  {title} {\bibinfo {title} {Probing topological quantum matter with scanning
  tunnelling microscopy},\ }\href {https://doi.org/10.1038/s42254-021-00293-7}
  {\bibfield  {journal} {\bibinfo  {journal} {Nat. Rev. Phys.}\ }\textbf
  {\bibinfo {volume} {3}},\ \bibinfo {pages} {249} (\bibinfo {year}
  {2021})}\BibitemShut {NoStop}%
\bibitem [{\citenamefont {Wiesendanger}(2009)}]{Wiesendanger2009}%
  \BibitemOpen
  \bibfield  {author} {\bibinfo {author} {\bibfnamefont {R.}~\bibnamefont
  {Wiesendanger}},\ }\bibfield  {title} {\bibinfo {title} {Spin mapping at the
  nanoscale and atomic scale},\ }\href
  {https://doi.org/10.1103/RevModPhys.81.1495} {\bibfield  {journal} {\bibinfo
  {journal} {Rev. Mod. Phys.}\ }\textbf {\bibinfo {volume} {81}},\ \bibinfo
  {pages} {1495} (\bibinfo {year} {2009})}\BibitemShut {NoStop}%
\bibitem [{\citenamefont {Yazyev}\ \emph {et~al.}(2010)\citenamefont {Yazyev},
  \citenamefont {Moore},\ and\ \citenamefont {Louie}}]{Yazyev2010}%
  \BibitemOpen
  \bibfield  {author} {\bibinfo {author} {\bibfnamefont {O.~V.}\ \bibnamefont
  {Yazyev}}, \bibinfo {author} {\bibfnamefont {J.~E.}\ \bibnamefont {Moore}},\
  and\ \bibinfo {author} {\bibfnamefont {S.~G.}\ \bibnamefont {Louie}},\
  }\bibfield  {title} {\bibinfo {title} {Spin polarization and transport of
  surface states in the topological insulators {Bi$_2$Se$_3$} and
  {Bi$_2$Te$_3$} from first principles},\ }\href
  {https://doi.org/10.1103/PhysRevLett.105.266806} {\bibfield  {journal}
  {\bibinfo  {journal} {Phys. Rev. Lett.}\ }\textbf {\bibinfo {volume} {105}},\
  \bibinfo {pages} {266806} (\bibinfo {year} {2010})}\BibitemShut {NoStop}%
\bibitem [{\citenamefont {Zhang}\ \emph {et~al.}(2013)\citenamefont {Zhang},
  \citenamefont {Liu},\ and\ \citenamefont {Zhang}}]{Zhang2013}%
  \BibitemOpen
  \bibfield  {author} {\bibinfo {author} {\bibfnamefont {H.}~\bibnamefont
  {Zhang}}, \bibinfo {author} {\bibfnamefont {C.-X.}\ \bibnamefont {Liu}},\
  and\ \bibinfo {author} {\bibfnamefont {S.-C.}\ \bibnamefont {Zhang}},\
  }\bibfield  {title} {\bibinfo {title} {Spin-orbital texture in topological
  insulators},\ }\href {https://doi.org/10.1103/PhysRevLett.111.066801}
  {\bibfield  {journal} {\bibinfo  {journal} {Phys. Rev. Lett.}\ }\textbf
  {\bibinfo {volume} {111}},\ \bibinfo {pages} {066801} (\bibinfo {year}
  {2013})}\BibitemShut {NoStop}%
\bibitem [{\citenamefont {Jozwiak}\ \emph {et~al.}(2013)\citenamefont {Jozwiak}
  \emph {et~al.}}]{Jozwiak2013}%
  \BibitemOpen
  \bibfield  {author} {\bibinfo {author} {\bibfnamefont {C.}~\bibnamefont
  {Jozwiak}} \emph {et~al.},\ }\bibfield  {title} {\bibinfo {title}
  {Photoelectron spin-flipping and texture manipulation in a topological
  insulator},\ }\href {https://doi.org/10.1038/nphys2572} {\bibfield  {journal}
  {\bibinfo  {journal} {Nat. Phys.}\ }\textbf {\bibinfo {volume} {9}},\
  \bibinfo {pages} {293} (\bibinfo {year} {2013})}\BibitemShut {NoStop}%
\bibitem [{\citenamefont {Xu}\ \emph {et~al.}(2012)\citenamefont {Xu} \emph
  {et~al.}}]{Xu2012}%
  \BibitemOpen
  \bibfield  {author} {\bibinfo {author} {\bibfnamefont {S.-Y.}\ \bibnamefont
  {Xu}} \emph {et~al.},\ }\bibfield  {title} {\bibinfo {title} {Hedgehog spin
  texture and {B}erry's phase tuning in a magnetic topological insulator},\
  }\href {https://doi.org/10.1038/nphys2351} {\bibfield  {journal} {\bibinfo
  {journal} {Nat. Phys.}\ }\textbf {\bibinfo {volume} {8}},\ \bibinfo {pages}
  {616} (\bibinfo {year} {2012})}\BibitemShut {NoStop}%
\bibitem [{\citenamefont {Fang}\ \emph {et~al.}(2012)\citenamefont {Fang},
  \citenamefont {Gilbert},\ and\ \citenamefont {Bernevig}}]{Fang2012}%
  \BibitemOpen
  \bibfield  {author} {\bibinfo {author} {\bibfnamefont {C.}~\bibnamefont
  {Fang}}, \bibinfo {author} {\bibfnamefont {M.~J.}\ \bibnamefont {Gilbert}},\
  and\ \bibinfo {author} {\bibfnamefont {B.~A.}\ \bibnamefont {Bernevig}},\
  }\bibfield  {title} {\bibinfo {title} {Bulk topological invariants in
  noninteracting point group symmetric insulators},\ }\href
  {https://doi.org/10.1103/PhysRevB.86.115112} {\bibfield  {journal} {\bibinfo
  {journal} {Phys. Rev. B}\ }\textbf {\bibinfo {volume} {86}},\ \bibinfo
  {pages} {115112} (\bibinfo {year} {2012})}\BibitemShut {NoStop}%
\bibitem [{SM()}]{SM}%
  \BibitemOpen
  \href@noop {} {}\bibinfo {note} {See the Supplemental Material below for
  derivations, models, numerical methods and tests, and a summary of the
  large-scale simulations.}\BibitemShut {Stop}%
\bibitem [{\citenamefont {Fu}(2009)}]{Fu2009}%
  \BibitemOpen
  \bibfield  {author} {\bibinfo {author} {\bibfnamefont {L.}~\bibnamefont
  {Fu}},\ }\bibfield  {title} {\bibinfo {title} {Hexagonal warping effects in
  the surface states of the topological insulator {Bi$_2$Te$_3$}},\ }\href
  {https://doi.org/10.1103/PhysRevLett.103.266801} {\bibfield  {journal}
  {\bibinfo  {journal} {Phys. Rev. Lett.}\ }\textbf {\bibinfo {volume} {103}},\
  \bibinfo {pages} {266801} (\bibinfo {year} {2009})}\BibitemShut {NoStop}%
\bibitem [{\citenamefont {Chen}(1990)}]{Chen1990}%
  \BibitemOpen
  \bibfield  {author} {\bibinfo {author} {\bibfnamefont {C.~J.}\ \bibnamefont
  {Chen}},\ }\bibfield  {title} {\bibinfo {title} {Tunneling matrix elements in
  three-dimensional space: The derivative rule and the sum rule},\ }\href
  {https://doi.org/10.1103/PhysRevB.42.8841} {\bibfield  {journal} {\bibinfo
  {journal} {Phys. Rev. B}\ }\textbf {\bibinfo {volume} {42}},\ \bibinfo
  {pages} {8841} (\bibinfo {year} {1990})}\BibitemShut {NoStop}%
\bibitem [{\citenamefont {St\v{r}eda}(1982)}]{Streda1982}%
  \BibitemOpen
  \bibfield  {author} {\bibinfo {author} {\bibfnamefont {P.}~\bibnamefont
  {St\v{r}eda}},\ }\bibfield  {title} {\bibinfo {title} {Theory of quantised
  {H}all conductivity in two dimensions},\ }\href
  {https://doi.org/10.1088/0022-3719/15/22/005} {\bibfield  {journal} {\bibinfo
   {journal} {J. Phys. C: Solid State Phys.}\ }\textbf {\bibinfo {volume}
  {15}},\ \bibinfo {pages} {L717} (\bibinfo {year} {1982})}\BibitemShut
  {NoStop}%
\bibitem [{\citenamefont {Otrokov}\ \emph {et~al.}(2019)\citenamefont {Otrokov}
  \emph {et~al.}}]{Otrokov2019}%
  \BibitemOpen
  \bibfield  {author} {\bibinfo {author} {\bibfnamefont {M.~M.}\ \bibnamefont
  {Otrokov}} \emph {et~al.},\ }\bibfield  {title} {\bibinfo {title} {Prediction
  and observation of an antiferromagnetic topological insulator},\ }\href
  {https://doi.org/10.1038/s41586-019-1840-9} {\bibfield  {journal} {\bibinfo
  {journal} {Nature}\ }\textbf {\bibinfo {volume} {576}},\ \bibinfo {pages}
  {416} (\bibinfo {year} {2019})}\BibitemShut {NoStop}%
\bibitem [{\citenamefont {Chang}\ \emph {et~al.}(2015)\citenamefont {Chang}
  \emph {et~al.}}]{Chang2015}%
  \BibitemOpen
  \bibfield  {author} {\bibinfo {author} {\bibfnamefont {C.-Z.}\ \bibnamefont
  {Chang}} \emph {et~al.},\ }\bibfield  {title} {\bibinfo {title}
  {High-precision realization of robust quantum anomalous {H}all state in a
  hard ferromagnetic topological insulator},\ }\href
  {https://doi.org/10.1038/nmat4204} {\bibfield  {journal} {\bibinfo  {journal}
  {Nat. Mater.}\ }\textbf {\bibinfo {volume} {14}},\ \bibinfo {pages} {473}
  (\bibinfo {year} {2015})}\BibitemShut {NoStop}%
\bibitem [{\citenamefont {Yin}\ \emph {et~al.}(2020)\citenamefont {Yin} \emph
  {et~al.}}]{Yin2020}%
  \BibitemOpen
  \bibfield  {author} {\bibinfo {author} {\bibfnamefont {J.-X.}\ \bibnamefont
  {Yin}} \emph {et~al.},\ }\bibfield  {title} {\bibinfo {title} {Quantum-limit
  {C}hern topological magnetism in {TbMn$_6$Sn$_6$}},\ }\href
  {https://doi.org/10.1038/s41586-020-2482-7} {\bibfield  {journal} {\bibinfo
  {journal} {Nature}\ }\textbf {\bibinfo {volume} {583}},\ \bibinfo {pages}
  {533} (\bibinfo {year} {2020})}\BibitemShut {NoStop}%
\bibitem [{\citenamefont {Ye}\ \emph {et~al.}(2018)\citenamefont {Ye} \emph
  {et~al.}}]{Ye2018}%
  \BibitemOpen
  \bibfield  {author} {\bibinfo {author} {\bibfnamefont {L.}~\bibnamefont {Ye}}
  \emph {et~al.},\ }\bibfield  {title} {\bibinfo {title} {Massive {D}irac
  fermions in a ferromagnetic kagome metal},\ }\href
  {https://doi.org/10.1038/nature25987} {\bibfield  {journal} {\bibinfo
  {journal} {Nature}\ }\textbf {\bibinfo {volume} {555}},\ \bibinfo {pages}
  {638} (\bibinfo {year} {2018})}\BibitemShut {NoStop}%
\bibitem [{\citenamefont {Tersoff}\ and\ \citenamefont
  {Hamann}(1985)}]{Tersoff1985}%
  \BibitemOpen
  \bibfield  {author} {\bibinfo {author} {\bibfnamefont {J.}~\bibnamefont
  {Tersoff}}\ and\ \bibinfo {author} {\bibfnamefont {D.~R.}\ \bibnamefont
  {Hamann}},\ }\bibfield  {title} {\bibinfo {title} {Theory of the scanning
  tunneling microscope},\ }\href {https://doi.org/10.1103/PhysRevB.31.805}
  {\bibfield  {journal} {\bibinfo  {journal} {Phys. Rev. B}\ }\textbf {\bibinfo
  {volume} {31}},\ \bibinfo {pages} {805} (\bibinfo {year} {1985})}\BibitemShut
  {NoStop}%
\bibitem [{\citenamefont {Wei{\ss}e}\ \emph {et~al.}(2006)\citenamefont
  {Wei{\ss}e}, \citenamefont {Wellein}, \citenamefont {Alvermann},\ and\
  \citenamefont {Fehske}}]{Weisse2006}%
  \BibitemOpen
  \bibfield  {author} {\bibinfo {author} {\bibfnamefont {A.}~\bibnamefont
  {Wei{\ss}e}}, \bibinfo {author} {\bibfnamefont {G.}~\bibnamefont {Wellein}},
  \bibinfo {author} {\bibfnamefont {A.}~\bibnamefont {Alvermann}},\ and\
  \bibinfo {author} {\bibfnamefont {H.}~\bibnamefont {Fehske}},\ }\bibfield
  {title} {\bibinfo {title} {The kernel polynomial method},\ }\href
  {https://doi.org/10.1103/RevModPhys.78.275} {\bibfield  {journal} {\bibinfo
  {journal} {Rev. Mod. Phys.}\ }\textbf {\bibinfo {volume} {78}},\ \bibinfo
  {pages} {275} (\bibinfo {year} {2006})}\BibitemShut {NoStop}%
\end{thebibliography}%

\clearpage
\let\addcontentsline\origaddcontentsline
\onecolumngrid
\setcounter{secnumdepth}{2}
\setcounter{section}{0}\setcounter{equation}{0}\setcounter{figure}{0}\setcounter{table}{0}
\renewcommand{\thesection}{S\arabic{section}}\renewcommand{\theequation}{S\arabic{equation}}
\renewcommand{\thefigure}{S\arabic{figure}}\renewcommand{\thetable}{S\arabic{table}}
\renewcommand{\theHsection}{S\arabic{section}}\renewcommand{\theHequation}{S\arabic{equation}}
\renewcommand{\theHfigure}{S\arabic{figure}}\renewcommand{\theHtable}{S\arabic{table}}
\begin{center}
{\large\bfseries Supplemental Material for ``Imaging the topological sign of a Dirac mass with spin-polarized tunneling''}\\[8pt]
Payman Kazemikhah, Ruolun Zhang, Byunghoon Kim, Xiangyu Luo, Zachary Ducorsky, Yu-Peng Zhu, Md. Shafayat Hossain, and M. Zahid Hasan\\[10pt]
\end{center}
\begin{quote}\small
This Supplemental Material gives the derivations behind the main text, the models and numerical methods, numerical tests of the analytic results, and a summary of the large-scale lattice simulations. Equations, figures, sections, and tables carry the prefix S. Unprefixed numbers refer to the main text.
\end{quote}

\tableofcontents

\section{Conventions and the physical mass sign}\label{S-sec:conventions}
\emph{Surface Hamiltonian.} Throughout, $\hbar=1$ unless units are stated. The surface band is
\begin{equation}
H(\mathbf k)=d_0(\mathbf k)+\mathbf d(\mathbf k)\cdot\bsig,\quad
\mathbf d=\bigl(-vk_y,\;vk_x,\;M-\beta k^2+\lambda\,\mathrm{Re}\,k_+^3\bigr),\quad d_0=Dk^2,
\label{S-eq:H}
\end{equation}
with $k_\pm=k_x\pm ik_y$ and $\mathrm{Re}\,k_+^3=k_x^3-3k_xk_y^2$. The Pauli matrices act on the Kramers doublet at $\bar\Gamma$ (the pseudospin, Sec.~\ref{S-sec:selection}). For a surface state built from $l_z=0$ orbitals only they act on the electron spin. The band energies are $E_\pm=d_0\pm|\mathbf d|$, and the upper band has Bloch vector $\mathbf n=\mathbf d/|\mathbf d|$ while the lower band has $-\mathbf n$. For $\beta=\lambda=D=0$ we write $E=\sqrt{v^2k^2+M^2}$.

\emph{Channels.} A tip effect is $W=g(\id+\mathbf P\cdot\bsig)$ with $|\mathbf P|\le1$, a point defect is $U=u_0+\mathbf u\cdot\bsig$, and the tunneling signal is $\rho_W(\mathbf r,E)=-\pi^{-1}\,\mathrm{Im}\Tr[WG^R(\mathbf r,\mathbf r;E)]=g[\rho_0+\mathbf P\cdot\bm\rho]$ with $\rho_a=-\pi^{-1}\,\mathrm{Im}\Tr[\sigma_aG^R]$. The tip-reversal asymmetry is $A=(\rho_{\mathbf P}-\rho_{-\mathbf P})/(\rho_{\mathbf P}+\rho_{-\mathbf P})=\mathbf P\cdot\bm\rho/\rho_0$.

\emph{Geometry.} For a band with projector $P(\mathbf k)$, the quantum geometric tensor is $Q_{ij}=\Tr[P\,\partial_iP\,\partial_jP]=\langle\partial_iu|(1-P)|\partial_ju\rangle$, the metric is $g=\mathrm{Re}\,Q$, and the Berry curvature is $\Omega=-2\,\mathrm{Im}\,Q_{xy}$. The coordinates $(x,y,z)$ are right handed with $\hat z$ the outward surface normal.

\thm{Berry curvature of the massive cone.} For a two-level band with Bloch vector $\mathbf n(\mathbf k)$, $g_{ij}=\tfrac14\partial_i\mathbf n\cdot\partial_j\mathbf n$ and $\Omega=-\tfrac12\,\mathbf n\cdot(\partial_x\mathbf n\times\partial_y\mathbf n)$. For Eq.~\eqref{S-eq:H} with $\beta=\lambda=0$ the occupied (lower) branch has
\begin{equation}
\Omega_{\rm occ}=+\frac{v^2M}{2E^3}
\label{S-eq:omega}
\end{equation}
for either sign of $v$.
The two-level formulas follow from $P=\tfrac12(\id+\mathbf n\cdot\bsig)$ and $\Tr[\sigma_a\sigma_b\sigma_c]=2i\epsilon_{abc}$. For the lower branch $\mathbf n=-\hat{\mathbf d}$, and $\hat{\mathbf d}\cdot(\partial_x\hat{\mathbf d}\times\partial_y\hat{\mathbf d})=\mathbf d\cdot(\partial_x\mathbf d\times\partial_y\mathbf d)/|\mathbf d|^3$. With $\partial_x\mathbf d=(0,v,0)$ and $\partial_y\mathbf d=(-v,0,0)$, the triple product is $v^2M$, which depends on $v$ only through $v^2$.

\emph{Mass sign.} We call $\sgn\Omega_{\rm occ}$ the mass sign or Hall sign. It is independent of the basis of the doublet, it equals the sign of the half-quantized surface Hall response [Eq.~\eqref{S-eq:streda}], and it fixes the circulation of domain-boundary modes (Sec.~\ref{S-sec:walls}). For Eq.~\eqref{S-eq:H}, $\sgn\Omega_{\rm occ}=\sgn M$. More generally, if $(d_x,d_y)$ winds once positively around $\bar\Gamma$, then $\sgn\Omega_{\rm occ}(\bar\Gamma)=\sgn d_z(\bar\Gamma)$.

\section{The sign group}\label{S-sec:sign}
\thm{Covariance of tunneling maps.} Let $\Theta=VK$ be antiunitary and on site, with $V$ a unitary acting on the internal indices and $K$ complex conjugation in the position basis, and let $H'=\Theta H\Theta^{-1}$. Then for every Hermitian effect $W$, position $\mathbf r$, and energy $E$,
\begin{equation}
\mathrm{Im}\Tr\bigl[W\,G^R_{H'}(\mathbf r,\mathbf r;E)\bigr]=\mathrm{Im}\Tr\bigl[(\Theta^{-1}W\Theta)\,G^R_{H}(\mathbf r,\mathbf r;E)\bigr].
\label{S-eq:covariance}
\end{equation}
For a unitary on-site $U$ and $H'=UHU^\dagger$, $\Tr[WG^R_{H'}]=\Tr[U^\dagger WUG^R_H]$.
From $\Theta(E+i0-H)\Theta^{-1}=E-i0-H'$ we get $G^R_{H'}=\Theta G^A_H\Theta^{-1}$. For antiunitary $\Theta$ and any operator $X$, $\Tr[\Theta X\Theta^{-1}]=\overline{\Tr X}$. Hence $\Tr[WG^R_{H'}]=\Tr[\Theta(\Theta^{-1}W\Theta)G^A_H\Theta^{-1}]=\overline{\Tr[W'G^A_H]}$ with $W'=\Theta^{-1}W\Theta$. Since $G^A=(G^R)^\dagger$ and $W'$ is Hermitian, $\Tr[W'G^A]=\overline{\Tr[W'G^R]}$. Combining, $\Tr[WG^R_{H'}(\mathbf r,\mathbf r)]=\Tr[W'G^R_H(\mathbf r,\mathbf r)]$, which is stronger than Eq.~\eqref{S-eq:covariance}. The position-diagonal block is preserved because $\Theta$ is on site. The unitary case is immediate.

The lemma holds for any Hamiltonian, including defect potentials, the exact $T$ matrix, disorder, and broadening, provided the image of every term is included in $H'$.

\thm{Action of the sign group.} Let $\Theta_\perp=\sigma_xK$, $\Theta_T=i\sigma_yK$, and $R_z=i\sigma_z$. Then $\Theta_\perp\sigma_{x,y}\Theta_\perp^{-1}=\sigma_{x,y}$, $\Theta_\perp\sigma_z\Theta_\perp^{-1}=-\sigma_z$, $\Theta_T\bsig\Theta_T^{-1}=-\bsig$, $R_z\sigma_{x,y}R_z^{-1}=-\sigma_{x,y}$, $R_z\sigma_zR_z^{-1}=\sigma_z$, and $K\mathbf kK=-\mathbf k$. The resulting action on the parameters of Eq.~\eqref{S-eq:H}, on the perpendicular field (orbital and Zeeman), and on the channels is given in Table~\ref{S-tab:action}. The three operations commute up to phases and, together with the identity, form a Klein four-group.

\begin{table}[b]
\caption{Images of the model parameters under the sign group. Entries give the factor multiplying each parameter. $B$ is the perpendicular field, entering through $\mathbf k\to\mathbf k+e\mathbf A$ and through a Zeeman term $\propto B\sigma_z$.}
\label{S-tab:action}
\begin{ruledtabular}
\begin{tabular}{lccccccccc}
& $v$ & $M$ & $\beta$ & $\lambda$ & $D$ & $B$ & $\mathbf P_\parallel,\mathbf u_\parallel$ & $P_z,u_z$ & $u_0$\\ \hline
$\Theta_\perp=\sigma_xK$ & $-$ & $-$ & $-$ & $+$ & $+$ & $-$ & $+$ & $-$ & $+$\\
$\Theta_T=i\sigma_yK$ & $+$ & $-$ & $-$ & $+$ & $+$ & $-$ & $-$ & $-$ & $+$\\
$R_z=i\sigma_z$ & $-$ & $+$ & $+$ & $+$ & $+$ & $+$ & $-$ & $+$ & $+$
\end{tabular}
\end{ruledtabular}
\end{table}

The Pauli relations follow from $\sigma_y^*=-\sigma_y$ and the anticommutation relations. Under $\Theta_\perp$ the kinetic term $v(k_x\sigma_y-k_y\sigma_x)$ becomes $v(-k_x\sigma_y+k_y\sigma_x)$, which is the kinetic term with $v\to-v$. The term $(M-\beta k^2)\sigma_z$ changes sign, $\mathrm{Re}\,k_+^3\sigma_z$ is invariant, because $k^3\to-k^3$ and $\sigma_z\to-\sigma_z$, and $Dk^2$ is invariant. With a vector potential, $\mathbf k+e\mathbf A\to-\mathbf k+e\mathbf A=-(\mathbf k-e\mathbf A)$ because $\mathbf A(\mathbf r)$ is real, so $\mathbf A\to-\mathbf A$. The Zeeman term $\propto B\sigma_z$ is reversed as well, consistently. Under $\Theta_T$ both $\mathbf k$ and $\bsig$ are reversed. The kinetic and warping terms are invariant, while $(M-\beta k^2)\sigma_z$ and every coupling to $B$ are reversed. Under $R_z$ only $\sigma_{x,y}$ are reversed, which is equivalent to $v\to-v$. The images of $W$ and $U$ follow from the Pauli relations. Finally, $\Theta_\perp\Theta_T=\sigma_x(i\sigma_y)^*K^2=\sigma_x(i\sigma_y)=-\sigma_z\propto R_z$, and similarly for the other products.

\emph{Lattice model F1.} Model F1 has $\mathbf d(\mathbf k)=(-t_1\sin k_y,\,t_1\sin k_x,\,m+t_2(\cos k_x+\cos k_y))$ (a two-band Chern-insulator lattice model written in topological-insulator form) with mass $M=m+2t_2$ at $\Gamma$. On the Bloch Hamiltonian, $\Theta H(\mathbf k)\Theta^{-1}=VH(-\mathbf k)^*V^\dagger$ for $\Theta=VK$. One finds $\Theta_\perp:\mathbf d\to-\mathbf d$, that is $(t_1,m,t_2)\to-(t_1,m,t_2)$, $\Theta_T:(t_1,m,t_2)\to(t_1,-m,-t_2)$, and $R_z:t_1\to-t_1$. Peierls phases are conjugated by $K$, so the flux per plaquette $\phi$ is reversed by both antiunitaries. The regulator $t_2(\cos k_x+\cos k_y)\sigma_z$ is therefore odd under time reversal. This is the origin of the lattice artefact of Sec.~\ref{S-sec:walls}.

\thm{Invariant signs.} Write the signs as $s=(s_v,s_M,s_B,s_\parallel,s_z)$. A monomial $\prod s_i^{a_i}$ with $a\in\mathbb Z_2^5$ is invariant if and only if $Ra=0$ over $\mathbb Z_2$, where the rows of $R$ are the reversal patterns $(1,1,1,0,1)$, $(0,1,1,1,1)$, and $(1,0,0,1,0)$ of Table~\ref{S-tab:action}. The third row is the sum of the first two, so $\mathrm{rank}\,R=2$ and the invariant subgroup has rank $5-2=3$. It is generated by $s_Ms_z$, $s_vs_Ms_\parallel$, and $s_Ms_B$, and its seven nontrivial elements are $s_Bs_z$, $s_Ms_z$, $s_Ms_B$, $s_vs_\parallel s_z$, $s_vs_Bs_\parallel$, $s_vs_Ms_\parallel$, and $s_vs_Ms_Bs_\parallel s_z$.
A monomial is invariant under a group element if and only if an even number of its factors are reversed by that element. The remaining statements are linear algebra over $\mathbb Z_2$.

By the covariance relation of Sec.~\ref{S-sec:sign}, two parameter sets related by a group element produce identical data once the channels are mapped accordingly. Tunneling data therefore determine at most the invariant signs. Conversely, each generator is determined by the corresponding calibrated reference: $\sgn(MP_z)$ by the sign of the on-axis band-edge asymmetry (the on-axis sign rule of Sec.~\ref{S-sec:selection}), $\sgn(MB)$ by the position of the zeroth Landau level (Sec.~\ref{S-sec:landau}), and $\sgn(vMP_\parallel)$ by in-plane spin-resolved quasiparticle interference (QPI) interpreted with an exact forward model, through the coherence factor below.

\thm{Born coherence factors.} For $\mathbf n'=(-n_x,-n_y,n_z)$, the Bloch vector at $-\mathbf k$ on the isotropic texture,
\begin{equation}
P(\mathbf n)P(\mathbf n')=\tfrac12\bigl[n_z^2+n_z\sigma_z+in_z(n_y\sigma_x-n_x\sigma_y)\bigr],
\label{S-eq:coherence}
\end{equation}
so that $\Tr[PP']=n_z^2$, $\Tr[\sigma_zPP']=n_z$, $\Tr[\sigma_xPP']=in_yn_z$, and $\Tr[\sigma_yPP']=-in_xn_z$. For the upper band $n_yn_z=vk_xM/E^2\propto vM$.
Expand $(\id+\mathbf n\cdot\bsig)(\id+\mathbf n'\cdot\bsig)=(1+\mathbf n\cdot\mathbf n')\id+(\mathbf n+\mathbf n')\cdot\bsig+i(\mathbf n\times\mathbf n')\cdot\bsig$ and insert $\mathbf n\cdot\mathbf n'=2n_z^2-1$, $\mathbf n+\mathbf n'=(0,0,2n_z)$, and $\mathbf n\times\mathbf n'=(2n_yn_z,-2n_xn_z,0)$.

\begin{figure}[t]
\includegraphics[width=\textwidth]{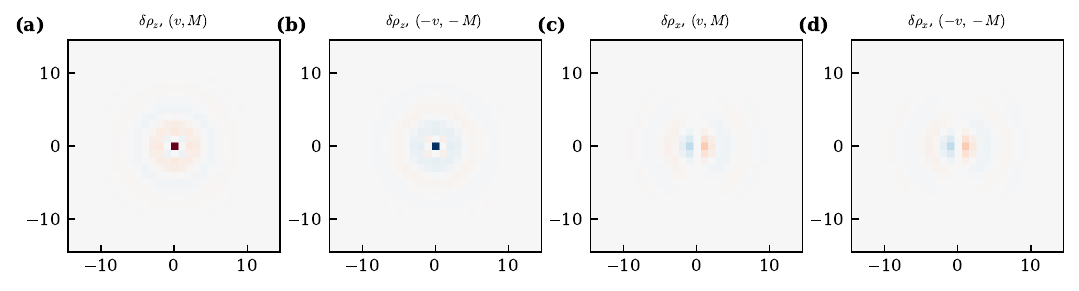}
\caption{Exact $T$-matrix maps around a scalar defect on model F1 ($E=0.6$, $\eta=0.05$, $64\times64$ torus). (a,b) The $z$-channel maps of $(v,M)$ and $(-v,-M)$ are exact negatives of each other (sum $4\times10^{-18}$). (c,d) The $x$-channel maps are identical (difference $4\times10^{-18}$). The signal scale is $6\times10^{-2}$.}
\label{S-fig:signmaps}
\end{figure}

\emph{Static maps carry no chirality.} Under $\Theta_\perp$ positions are unchanged, and by the covariance relation of Sec.~\ref{S-sec:sign} every map taken with unpolarized or in-plane-polarized tips and scalar or in-plane defects is identical for $(v,M)$ and $(-v,-M)$. No real-space pattern can therefore reveal the Hall sign without a time-reversal-odd reference of known sign, however chiral it looks. The exact $T$-matrix maps of F1 confirm this to $2\times10^{-17}$ (Fig.~\ref{S-fig:signmaps}).

\emph{Magnetic defects as references.} A defect moment $u_z$ of known sign is itself a time-reversal-odd reference: $\sgn(Mu_z)$ is invariant (the invariant signs of Sec.~\ref{S-sec:sign}), so in-plane tips can then determine the mass sign. Exact $T$-matrix maps show both cases (Sec.~\ref{S-sec:campaigns}). With in-plane tips and a $z$-moment defect held fixed, the maps of $(v,M)$ and $(-v,-M)$ differ. When the defect moment is imaged as well ($u_z\to-u_z$, a moment of unknown sign), they coincide to $2.5\times10^{-14}$ in all 360 settings.

\section{Kramers null theorem and field reversal}\label{S-sec:kramers}
\thm{Kramers null result.} If $\Theta_TH\Theta_T^{-1}=H$, with $\Theta_T=(\id_{\rm orb}\otimes i\sigma_y)K$ acting on site, then $\Tr_{\rm orb}[\sigma_aG^R(\mathbf R,\mathbf R;E)]=0$ for $a=x,y,z$, every site $\mathbf R$, and every complex energy in the upper half plane. The same holds with $\sigma_a$ replaced by $F\otimes\sigma_a$ for any real symmetric orbital filter $F$.
This follows from the covariance relation of Sec.~\ref{S-sec:sign} with $H'=H$ and $W=\sigma_a$, for which $\Theta_T^{-1}\sigma_a\Theta_T=-\sigma_a$: $\Tr[\sigma_aG^R(\mathbf R,\mathbf R)]=\Tr[(-\sigma_a)G^R(\mathbf R,\mathbf R)]$. For $F\otimes\sigma_a$ with $F$ real symmetric, $K$ leaves $F$ invariant.

The theorem covers arbitrary disorder, orbitals, spin--orbit coupling, warping, and nonmagnetic defects. For interacting time-reversal-invariant systems the same conclusion holds for the exact single-particle Green function, whose Lehmann representation inherits the antiunitary symmetry. A random multi-orbital test gives $6\times10^{-15}$, against $0.24$ once a random exchange field is added.

\thm{Field reversal.} Let $H(\mathbf m,B)$ contain an arbitrary magnetic texture $\mathbf m(\mathbf r)$ and a perpendicular field $B$, entering through Peierls phases and Zeeman terms, with a time-reversal-invariant remainder, so that $\Theta_TH(\mathbf m,B)\Theta_T^{-1}=H(-\mathbf m,-B)$. Then $\rho_z(\mathbf r;\mathbf m,B)=-\rho_z(\mathbf r;-\mathbf m,-B)$ site by site, and the field-averaged signal $\bar\rho_z(\mathbf m)=\tfrac12[\rho_z(\mathbf m,B)+\rho_z(\mathbf m,-B)]$ is exactly odd, $\bar\rho_z(-\mathbf m)=-\bar\rho_z(\mathbf m)$, while $\rho_0$ is even.
the covariance relation of Sec.~\ref{S-sec:sign} with $W=\sigma_z$ gives the first statement. Averaging over $\pm B$ gives the second.

The statement holds to all orders in $B$, including Landau quantization, and in the presence of disorder (a numerical test on a disordered $10\times10$ cluster with flux and Zeeman terms gives a residual of $2\times10^{-16}$). It requires that neither the sample magnetization nor the tip polarization switches between $+B$ and $-B$, which calls for fields below the coercive field and for remanent or antiferromagnetic tips. Tip-induced fields reverse with the tip polarization and not with $B$, so they are not removed by the average (Sec.~\ref{S-sec:tip}).

\section{Local spectroscopy}\label{S-sec:local}
\thm{Contour-average identity.} For a two-band surface, $G(\mathbf k,z)=\sum_{s=\pm}P_s(\mathbf k)/(z-E_s(\mathbf k))$ with $P_\pm=\tfrac12(\id\pm\mathbf n\cdot\bsig)$ and $\Tr[\sigma_aP_\pm]=\pm n_a$. Therefore, at energies where only branch $s$ contributes,
\begin{equation}
\frac{\rho_a(E)}{\rho_0(E)}=\frac{\oint_{E_s=E}\frac{d\ell}{|\nabla E_s|}\,s\,n_a}{\oint_{E_s=E}\frac{d\ell}{|\nabla E_s|}}\equiv\langle s\,n_a\rangle_E .
\label{S-eq:contour}
\end{equation}
Take $-\pi^{-1}\mathrm{Im}$ of the trace, integrate over $\mathbf k$, and use $\int d^2k\,\delta(E-E_s)f=\oint d\ell\,f/|\nabla E_s|$.

\thm{Isotropic cone.} For $\beta=\lambda=D=0$, $\rho_0=|E|\Theta(|E|-|M|)/2\pi v^2$ and $\rho_z=(M/2\pi v^2)[\Theta(E-|M|)-\Theta(-E-|M|)]$, so that $A=\rho_z/\rho_0=M/E$ for signed $E$ with $|E|>|M|$. With Lorentzian broadening $L_\eta(x)=\eta/\pi(x^2+\eta^2)$,
\begin{align}
\rho_z(E)&=\frac{M}{2\pi^2v^2}\left[\arctan\frac{E-|M|}{\eta}+\arctan\frac{E+|M|}{\eta}\right],\label{S-eq:rhoz}\\
\rho_0(E)&=\frac{1}{2\pi v^2}\Bigl[\mathcal F(\Lambda)-\mathcal F(|M|)\Bigr],\quad
\mathcal F(\epsilon)=\frac{\eta}{2\pi}\ln\bigl[((\epsilon-E)^2+\eta^2)((\epsilon+E)^2+\eta^2)\bigr]+\frac{E}{\pi}\Bigl[\arctan\frac{\epsilon-E}{\eta}-\arctan\frac{\epsilon+E}{\eta}\Bigr],\label{S-eq:rho0}
\end{align}
with an energy cutoff $\Lambda$. The numerator is cutoff free. The denominator depends on $\Lambda$ only through the term $\eta\ln\Lambda$.
The upper band contributes spin $n_z=M/\epsilon$ with density $\epsilon/2\pi v^2$ at energy $\epsilon$, and the lower band contributes $-M/\epsilon$ at $-\epsilon$. Hence $\rho_z=(M/2\pi v^2)\int_{|M|}^\infty d\epsilon\,[L_\eta(E-\epsilon)-L_\eta(E+\epsilon)]$, which evaluates to Eq.~\eqref{S-eq:rhoz}. Equation~\eqref{S-eq:rho0} follows from $\int(u+E)L_\eta(u)\,du=\frac{\eta}{2\pi}\ln(u^2+\eta^2)+\frac E\pi\arctan\frac u\eta$.

\begin{figure}[t]
\includegraphics[width=\textwidth]{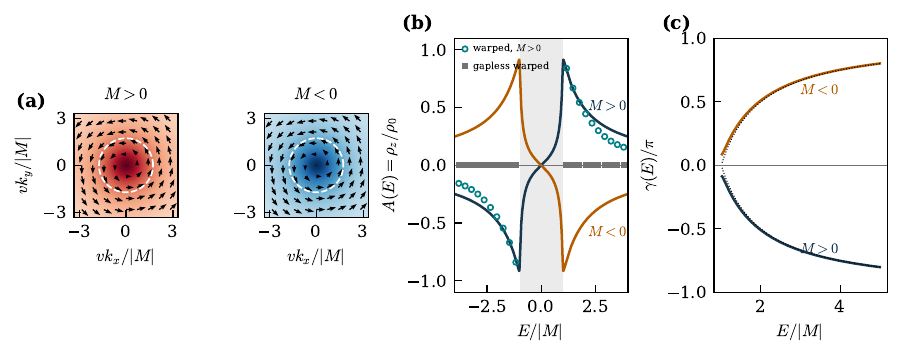}
\caption{Local estimator. (a)~Upper-band pseudospin texture (arrows: in-plane part, color: $n_z$ from $-1$, blue, to $+1$, red). Dashed contours at $E=2|M|$. (b)~$A=\rho_z/\rho_0$ for $M=\pm0.3$ ($v=1$) from Eqs.~\eqref{S-eq:rhoz} and \eqref{S-eq:rho0} with $\eta=0.004$. Circles: warped cone ($\lambda=0.6$). Squares: gapless warped cone, identically zero. (c)~Contour Berry phase reconstructed from $A(E)$ with $\kappa=1$ (solid) and exact (dotted). Opposite domains have equal metric and opposite phase.}
\label{S-fig:local}
\end{figure}

\emph{Band-edge value.} At $E=|M|$, $\rho_z=(M/2\pi^2v^2)\arctan(2|M|/\eta)\to M/4\pi v^2$ and $\rho_0\to|M|/4\pi v^2$, so $A(\pm|M|)\to\pm\sgn M$ [Fig.~\ref{S-fig:local}(b)]. For $\eta/|M|=1/30$, $1/300$, and $1/3000$ the broadened values are $0.81$, $0.973$, and $0.997$. The numerical polar-grid integration agrees with Eqs.~\eqref{S-eq:rhoz} and \eqref{S-eq:rho0} to $6\times10^{-5}$ when the same cutoff is used.

\thm{Symmetries of the local signal.} In the translation-invariant continuum, $\rho_z(M,\beta,\lambda)=-\rho_z(-M,-\beta,\lambda)=\rho_z(M,\beta,-\lambda)$. In particular $\rho_z\equiv0$ on a gapless surface with arbitrary warping.
The first equality is the action of the sign group (Sec.~\ref{S-sec:sign}) for $\Theta_T$ with the covariance relation of Sec.~\ref{S-sec:sign}. For the second, the joint rotation by $\pi/3$ of $\mathbf k$ and of the pseudospin about $\hat z$ leaves $v(\mathbf k\times\bsig)_z$, $\sigma_z$, $k^2$, and $M$ invariant and maps $\mathrm{Re}\,k_+^3\to-\mathrm{Re}\,k_+^3$. Local densities of a translation-invariant system do not depend on position, and $\sigma_z$ is invariant, so $\rho_z(\lambda)=\rho_z(-\lambda)$.

Numerically the residuals are $2\times10^{-17}$ for both equalities and $10^{-18}$ for the gapless warped surface. The warped contour average at $E=2|M|$ and $\lambda=0.6$ is $0.4626$ ($\eta\to0$), against $0.462$ from the broadened calculation.

\emph{Band curvature.} A single surface that is time-reversal invariant at $M=0$ cannot contain $\beta k^2\sigma_z$, because that term is odd under $\Theta_T$. Exchange can generate $\beta\propto J$, so only joint oddness in $(M,\beta)$ is physical. Lattice regularizations with a fixed regulator produce a sign change of $\langle n_z\rangle_E$ at $k^2=M/\beta$ in one domain only. On a single surface this is an artefact of the regularization (Sec.~\ref{S-sec:walls}). The same term is TR even when it multiplies a surface-index matrix in a thin-film (hybridized) model, where it describes a different physical effect.

\emph{Berry-phase spectroscopy.} By Stokes' theorem the Berry phase of the counterclockwise contour of the upper band at energy $E$ is $\gamma(E)=\int_{E_+<E}\Omega_{\rm upper}\,d^2k$. For the isotropic cone, $\int_0^{k_E}(-v^2M/2E^3)\,2\pi k\,dk=-\pi\,\sgn M\,(1-|M|/E)$, so
\begin{equation}
\gamma(E)=-\pi\bigl[1-\langle n_z\rangle_E\bigr]\pmod{2\pi},\qquad \gamma(E)=-\pi\bigl[1-A(E)/\kappa\bigr],
\label{S-eq:berryphase}
\end{equation}
where $\kappa$ is the effective tip polarization of Sec.~\ref{S-sec:selection}. The magnitude $|\gamma|=\pi(1-|n_z|)$ is half the solid angle of the cap enclosed by the contour on the Bloch sphere. A Wilson loop reproduces Eq.~\eqref{S-eq:berryphase} to $6\times10^{-6}$, and Fig.~\ref{S-fig:local}(c) shows the reconstruction from $A(E)$. With warping and $M\ne0$ the phase changes at second order in $\lambda$. At $M=0$, $\gamma=\pi$ exactly for any warping, because the texture is antipodally symmetric, $\mathbf n(-\mathbf k)=(-n_x,-n_y,n_z)$ with $n_z$ odd.

\section{Orbital filters as Lorentz boosts}\label{S-sec:boost}
Let the surface doublet $\{|+\rangle,|-\rangle\}$ (pseudospin $\tau$) be probed by a tip whose spin-resolved amplitudes are $\psi_\sigma=\sum_{j=\pm}F_{\sigma j}c_j$ for a state $c_+|+\rangle+c_-|-\rangle$. The measured (unnormalized) spin density matrix is $F\rho_\tau F^\dagger$. Its four-vector is therefore a linear image of the pseudospin four-vector $(1,\mathbf n)$ under an element of $\mathbb R_+\times\mathrm{SO}^+(3,1)$ (for $\det F\ne0$), that is, a Lorentz transformation of the Bloch four-vector.

\thm{Diagonal filters are $z$ boosts.} For $F=\mathrm{diag}(a_+,a_-)$ with $w_\pm=|a_\pm|^2$ and $\delta=(w_+-w_-)/(w_++w_-)$, the measured Bloch vector is
\begin{equation}
r_z=\frac{n_z+\delta}{1+\delta n_z},\qquad \mathbf r_\perp=\frac{\sqrt{1-\delta^2}}{1+\delta n_z}\,\mathcal R_\chi\,\mathbf n_\perp,
\label{S-eq:boost}
\end{equation}
where $\mathcal R_\chi$ is an in-plane rotation by $\chi=\arg(a_+\bar a_-)$. This is a Lorentz boost along $z$ with rapidity $\zeta=\ln|a_+/a_-|$, $\delta=\tanh\zeta$. The poles $\mathbf n=\pm\hat z$ are fixed points.
$F\rho F^\dagger$ has diagonal entries $w_\pm(1\pm n_z)/2$ and off-diagonal entry $a_+\bar a_-(n_x-in_y)/2$. Normalizing by the trace $[(w_++w_-)+(w_+-w_-)n_z]/2$ gives Eq.~\eqref{S-eq:boost}.

On a $C_n$ axis the tip filter is diagonal in the $j_z$ basis (the axial spin purity of Sec.~\ref{S-sec:selection} below), so on-axis filtering is a $z$ boost and cannot move the band-edge poles. Off axis, $F$ is a general $2\times2$ matrix, the Lorentz transformation includes rotations, and the poles move.

\section{The on-axis selection rule}\label{S-sec:selection}
\thm{Spin purity on rotation axes.} Let the surface (including the magnetization, taken along $\hat z$) be invariant under a rotation $C_n$ ($n\ge2$) about an axis $\ell$ parallel to $\hat z$. Let $\psi$ be a Bloch state at $\bar\Gamma$ with $C_n\psi=e^{-2\pi ij_z/n}\psi$ and $j_z\equiv\pm\frac12\pmod n$. Then for every $\mathbf r\in\ell$ (inside or outside the crystal), $\psi_\downarrow(\mathbf r)=0$ if $j_z=+\frac12$ and $\psi_\uparrow(\mathbf r)=0$ if $j_z=-\frac12$. The same holds for $(D\psi)(\mathbf r)$ for every linear differential operator $D$ commuting with rotations about $\ell$. The conclusion is the same for every $C_n$ axis of the surface.
The rotation acts as $(C_n\psi)(\mathbf r)=e^{-i\pi\sigma_z/n}\psi(\mathcal R^{-1}\mathbf r)$, with $\mathcal R$ the spatial rotation about $\ell$. For $\mathbf r\in\ell$, $\mathcal R^{-1}\mathbf r=\mathbf r$, and the eigenvalue equation becomes $e^{-i\pi s/n}\psi_s(\mathbf r)=e^{-2\pi ij_z/n}\psi_s(\mathbf r)$ with $s=\pm1$. For $j_z=+\frac12$ and $s=-1$ this requires $e^{i\pi/n}=e^{-i\pi/n}$, which is false for $n\ge2$, so $\psi_\downarrow(\mathbf r)=0$. The case $j_z=-\frac12$ is analogous. If $[D,C_n]=0$, then $D\psi$ has the same eigenvalue and the argument applies to it. Two $C_n$ axes of the surface differ by a lattice translation $T_{\mathbf a}$, so $C_n'=T_{\mathbf a}C_nT_{\mathbf a}^{-1}$ up to a translation, and translations act as the identity at $\bar\Gamma$.

In Chen's derivative rule~\cite{Chen1990} an $s$, $p_z$, or $d_{z^2}$ tip couples through $\psi$, $\partial_z\psi$, or $(\partial_z^2-\nabla^2/3)\psi$, respectively, all of which commute with rotations about the axis. the axial spin purity of Sec.~\ref{S-sec:selection} thus gives $c_z^{\rm eff}=1$ for such tips on atop and hollow sites. In the Tersoff--Hamann picture~\cite{Tersoff1985} with an $s$ tip, the spin-resolved signal is $\propto|\psi_\sigma(\mathbf r_{\rm tip})|^2$ directly.

\thm{Spin purity of the $\mathbf G=0$ component.} The laterally averaged spinor $\bar\psi(z)=A_{\rm cell}^{-1}\int_{\rm cell}\psi(\mathbf r_\parallel,z)\,d^2r_\parallel$ of a band-edge state with $j_z=\pm\frac12$ is spin pure. In vacuum, $\psi(\mathbf r)=\sum_{\mathbf G}c_{\mathbf G}e^{i\mathbf G\cdot\mathbf r_\parallel}e^{-\kappa_{\mathbf G}z}$ with $\kappa_{\mathbf G}=\sqrt{\kappa^2+|\mathbf G|^2}$ at $\bar\Gamma$. Hence at every lateral position $1-|c_z^{\rm eff}|=O(e^{-2(\kappa_{\mathbf G_1}-\kappa)z})$, with $\mathbf G_1$ the shortest nonzero reciprocal vector.
Averaging over the cell commutes with rotation about any $C_n$ axis, because lateral averages of periodic functions are rotation invariant. The eigenvalue argument of the axial spin purity of Sec.~\ref{S-sec:selection} then applies to $\bar\psi$. The spin-minority amplitude at a generic position is carried only by $\mathbf G\ne0$ components, whose relative weight decays as stated.

For Bi$_2$Se$_3$ ($a=4.14$~\AA, $|\mathbf G_1|=4\pi/\sqrt3a=1.75$~\AA$^{-1}$) and a vacuum decay constant $\kappa=0.512\sqrt{\phi[\mathrm{eV}]}$~\AA$^{-1}\approx1.1$~\AA$^{-1}$, $\kappa_{\mathbf G_1}-\kappa\approx0.95$~\AA$^{-1}$. The corrugation of $1-|c_z^{\rm eff}|$ is therefore $\sim e^{-9.5}\approx10^{-4}$ at $z=5$~\AA. Tip orbitals with $m\ne0$ couple only to $\mathbf G\ne0$ components at $\bar\Gamma$. On the axis they see the \emph{minority} spin exactly. In the model below a pure $p_x$ tip gives $-1$ times the $s$-tip value. A tip's $m\ne0$ content therefore reduces $c_z^{\rm eff}$ on the axis by a weight that is suppressed as $e^{-2\Delta\kappa z}$. A band-edge state with $j_z\equiv\frac32$ has no $\mathbf G=0$ component at all. It is dark in vacuum tunneling.

\thm{Threefold symmetry fixes the winding.} Let $n\ge3$, and let $H_{\rm eff}(\mathbf k)$ be the $2\times2$ Hamiltonian of the doublet in the basis $(|{+}\tfrac12\rangle,|{-}\tfrac12\rangle)$. Then the term linear in $\mathbf k$ of the off-diagonal element is $h_{+-}=b\,k_-$ for some complex $b$. Writing $H_{\rm eff}=d_0+\mathbf d\cdot\bm\tau$, the in-plane vector $(d_x,d_y)$ winds by $+1$ around $\bar\Gamma$ whenever $b\ne0$.
Rotation invariance means $UH_{\rm eff}(\mathbf k)U^\dagger=H_{\rm eff}(\mathcal R\mathbf k)$ with $U=\mathrm{diag}(e^{-i\pi/n},e^{i\pi/n})$, so $h_{+-}(\mathcal R\mathbf k)=e^{-2\pi i/n}h_{+-}(\mathbf k)$. Since $k_\pm\to e^{\pm2\pi i/n}k_\pm$, a term $a\,k_++b\,k_-$ must satisfy $a\,e^{2\pi i/n}=a\,e^{-2\pi i/n}$, which forces $a=0$ for $n\ge3$. Then $d_x-id_y=b(k_x-ik_y)$, so $(d_x,d_y)$ is $|b|(k_x,k_y)$ rotated by $\arg b$, which has winding $+1$.

For $n=2$ both $k_+$ and $k_-$ are allowed and the winding is not fixed by symmetry. the axial spin purity of Sec.~\ref{S-sec:selection} still holds, but the on-axis sign rule of Sec.~\ref{S-sec:selection} needs $n\ge3$.

\thm{The on-axis sign equals the Hall sign.} Under the assumptions of the axial spin purity of Sec.~\ref{S-sec:selection} with $n\ge3$, let the doublet disperse linearly ($b\ne0$), let the gap $2|\tilde M|$ be opened by $H_{\rm eff}(\bar\Gamma)=\tilde M\tau_z$, and let the band extrema lie at $\bar\Gamma$. Then the upper band-edge state has $j_z=\frac12\sgn\tilde M$, $\sgn\Omega_{\rm occ}=\sgn\tilde M$, and an axial tip on any $C_n$ axis sees the upper edge with spin $\sgn\Omega_{\rm occ}$ and the lower edge with spin $-\sgn\Omega_{\rm occ}$ [\Meq{axis}].
By the winding argument of Sec.~\ref{S-sec:selection} and the Berry-curvature formula of Sec.~\ref{S-sec:conventions} (with the winding $+1$ of the winding argument of Sec.~\ref{S-sec:selection}), $\sgn\Omega_{\rm occ}(\bar\Gamma)=\sgn d_z(\bar\Gamma)=\sgn\tilde M$. The upper edge state is the $\tau_z=\sgn\tilde M$ eigenstate, that is $j_z=\frac12\sgn\tilde M$. By the axial spin purity of Sec.~\ref{S-sec:selection} its spinor on the axis is pure with $s_z=2j_z$.

the on-axis sign rule of Sec.~\ref{S-sec:selection} was tested for 293 random parameter sets of the model below. In every case the on-axis sign equals $\sgn\Omega_{\rm occ}$, computed independently from the band projector, to machine precision on atop, hcp, and fcc sites, while the orbital-traced spin has the opposite sign in 161 cases.

\begin{figure}[t]
\includegraphics[width=\textwidth]{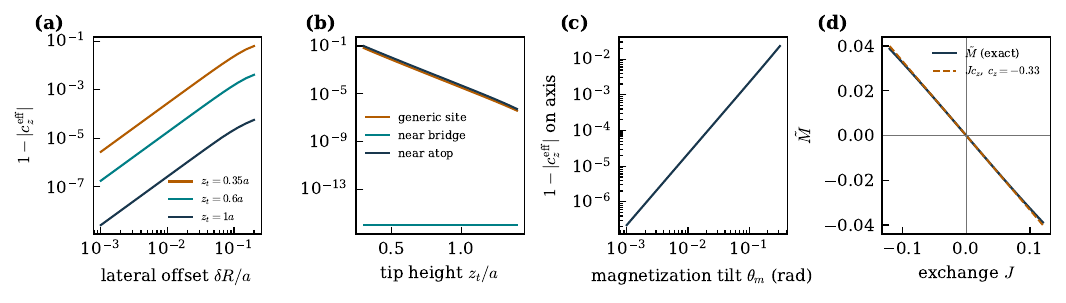}
\caption{Robustness of the selection rule in the model of Eq.~\eqref{S-eq:pmodel}. (a) $1-|c_z^{\rm eff}|$ versus lateral offset from the atop axis for three tip heights. Quadratic. (b) Versus tip height at three lateral positions. Exponential. The bridge site of the single-layer model is a $C_2$ axis and stays exact. (c) Versus tilt of the magnetization. Quadratic. (d) Exchange mass versus $J$ compared with the first-order result $Jc_z$ ($c_z=-1/3$).}
\label{S-fig:robust}
\end{figure}

\thm{Exchange mass.} Let $|{\pm}\rangle$ be the Kramers doublet at $J=0$, with $c_z=\langle{+}|\id_{\rm orb}\otimes\sigma_z|{+}\rangle$ the orbital-traced polarization of the $j_z=+\frac12$ partner. For an exchange coupling $J\,\id_{\rm orb}\otimes\sigma_z$, $\tilde M=Jc_z+O(J^2)$. The orbital-traced spin of the upper band edge is $\sgn(J)|c_z|+O(J)$, whereas the on-axis spin is $\sgn(Jc_z)$, so the ratio of the two signs is $\sgn c_z$.
Degenerate perturbation theory: $\langle\pm|\sigma_z|\pm\rangle=\pm c_z$ by time reversal, and $\langle+|\sigma_z|-\rangle=0$ by $j_z$ conservation (mod $n$). The upper edge is $|\sgn(\tilde M)\rangle$, whose orbital-traced spin is $\sgn(\tilde M)c_z=\sgn(J)|c_z|$.
Numerically $\tilde M/Jc_z=0.9999$ at $J=0.005$ and $0.9991$ at $J=0.02$ (Fig.~\ref{S-fig:robust}d).

\emph{Away from the band edge.} States at $\mathbf k\ne0$ are superpositions $c_+|{+}\rangle+c_-|{-}\rangle$ plus $O(ka)$ admixtures of remote bands. On axis, or in the $\mathbf G=0$ channel, the tip weighs $|{\pm}\rangle$ with $w_\pm$. By the boost form of Sec.~\ref{S-sec:boost} the measured asymmetry is the boost form, \Meq{boost}, with $\delta=(w_+-w_-)/(w_++w_-)$. Time reversal gives $w_+=w_-$ at $J=0$, so $\delta=O(J/\Delta_r)$ is odd in the magnetization, like the signal. Because the poles are fixed, the band-edge values remain exact. Far above the edge, where $|n_z|<|\delta|$, the sign of $A$ may follow $\delta$ instead of $n_z$. Sign sign assignments should therefore use the band-edge window $|E|\lesssim|\tilde M|/|\delta|$ or fit the boost form. In the model of Fig.~\ref{fig:selection}, $w_+=0.262$, $w_-=0.419$, and $\delta=-0.23$. This exchange is deliberately large. For realistic ratios $J/\Delta_r\sim10^{-2}$, $\delta$ is of the same order.

\emph{Model.} The $C_{3v}$ spin--orbital model used in \Mfig{selection} and Fig.~\ref{S-fig:robust} is a triangular lattice (constant $a=1$) with $(p_x,p_y,p_z)\otimes$ spin orbitals and Bloch Hamiltonian
\begin{equation}
H(\mathbf k)=\sum_{\bm\delta}T(\bm\delta)e^{i\mathbf k\cdot\bm\delta}\otimes\id+\mathrm{diag}(\epsilon_{xy},\epsilon_{xy},\epsilon_z)\otimes\id+\tfrac{\lambda_{\rm so}}{2}\mathbf L\cdot\bsig+J\,\id\otimes(\hat{\mathbf m}\cdot\bsig),
\label{S-eq:pmodel}
\end{equation}
with the six nearest-neighbor bonds $\bm\delta$, Slater--Koster hopping $T_{\alpha\beta}=V_\sigma\delta_\alpha\delta_\beta+V_\pi(\delta_{\alpha\beta}-\delta_\alpha\delta_\beta)$, and an orbital Rashba term $T_{z\alpha}=-T_{\alpha z}=\gamma\delta_\alpha$ ($\alpha=x,y$), which breaks $z\to-z$. Here $(L_k)_{ij}=-i\epsilon_{kij}$. Default parameters are $\epsilon_{xy}=-0.6$, $\epsilon_z=0.6$, $V_\sigma=0.3$, $V_\pi=-0.1$, $\lambda_{\rm so}=0.6$, $\gamma=0.3$, $J=0.08$, and $\hat{\mathbf m}=\hat z$. At $\bar\Gamma$ the $j_z=\pm\frac12$ block in the basis $(p_z s,\;p_{\pm}\bar s)$ is
\begin{equation}
\begin{pmatrix}E_z\pm J&\lambda_{\rm so}/\sqrt2\\ \lambda_{\rm so}/\sqrt2&E_{xy}-\lambda_{\rm so}/2\mp J\end{pmatrix},\qquad E_z=\epsilon_z+6V_\pi,\quad E_{xy}=\epsilon_{xy}+3(V_\sigma+V_\pi).
\label{S-eq:block}
\end{equation}
The lowest doublet has $c_z=-1/3$ at $J=0$. The $s$-wave tip at $\mathbf R=(x,y,z_t)$ has amplitudes $t_\alpha=e^{-\kappa|\mathbf R-\mathbf r_j|}(\mathbf R-\mathbf r_j)_\alpha/|\mathbf R-\mathbf r_j|$ (the $s$--$p\sigma$ overlap), with $\kappa=2.5/a$, summed over sites within $4.5a$. A single triangular layer has additional twofold axes through the bridge sites. By the axial spin purity of Sec.~\ref{S-sec:selection} these also give exact zeros [\Mfig{selection}(a)], whereas on a real $C_{3v}$ surface only the $C_3$ axes remain. The random parameter sets draw $\epsilon_{xy}\in[-1,0.2]$, $\epsilon_z\in[0.2,1]$, $V_\sigma\in[0.1,0.5]$, $V_\pi\in[-0.2,0]$, $\lambda_{\rm so}\in[0.2,1]$, $|\gamma|\in[0.1,0.5]$, and $|J|\in[0.02,0.12]$ uniformly, with random signs of $\gamma$ and $J$.

\emph{Robustness} (Fig.~\ref{S-fig:robust}). The deviation $1-|c_z^{\rm eff}|$ grows as $(\delta R)^2$ with the lateral offset (fitted exponent $1.94$), because the minority amplitude is linear in the offset. It grows as $\theta_m^2$ with the tilt of the magnetization (exponent $2.00$), which breaks $C_3$ at first order in amplitude. With height it decays as $e^{-2\Delta\kappa z}$, with a fitted rate $11.0/a$ against $10.4/a$ from the $\mathbf G=0$ purity of Sec.~\ref{S-sec:selection} for this tip model. Uniform strain leaves the purity intact in this model. A uniformly strained Bravais layer keeps its in-plane inversion, and Slater--Koster couplings of $p$ orbitals are even in the bond vector, so the $\bar\Gamma$ Hamiltonian keeps a twofold symmetry. Breaking $C_3$ in the Hamiltonian by a lateral shift $u$ of a subsurface layer (bilayer model) gives $1-|c_z^{\rm eff}|\approx0.07\,(u/a)^2$ (exponent $2.00$), independent of tip height, because the minority component then resides in the $p_z$ orbitals themselves.

\section{Landau quantization}\label{S-sec:landau}
With electron charge $-e$ and $\bm\pi=\mathbf p+e\mathbf A$, $[\pi_x,\pi_y]=-i\hbar eB$. For $eB>0$ the operator $a=(\pi_x-i\pi_y)/\sqrt{2\hbar eB}$ obeys $[a,a^\dagger]=1$. For $eB<0$ the roles of $\pi_x\mp i\pi_y$ are exchanged. With $\pi_x\sigma_y-\pi_y\sigma_x=\begin{pmatrix}0&-i(\pi_x-i\pi_y)\\ i(\pi_x+i\pi_y)&0\end{pmatrix}$, Eq.~\eqref{S-eq:H} with $\lambda=D=0$ becomes, for $eB>0$,
\begin{equation}
H=\begin{pmatrix}M-\beta\pi^2&-iv\sqrt{2\hbar eB}\,a\\ iv\sqrt{2\hbar eB}\,a^\dagger&-(M-\beta\pi^2)\end{pmatrix},\qquad \pi^2=\hbar|eB|(2a^\dagger a+1).
\label{S-eq:HLL}
\end{equation}
The spinor $(0,|0\rangle)$ is an exact eigenstate with
\begin{equation}
E_0=-\sgn(eB)\bigl(M+e_Z-\beta|eB|/\hbar\bigr),
\label{S-eq:E0}
\end{equation}
where a Zeeman term $e_Z\sigma_z$ has been included. The result is independent of $v$. In the sector $(|n-1\rangle,|n\rangle)$ the $2\times2$ problem gives $E_{n,\pm}=\pm[M_n^2+2n\hbar v^2|eB|]^{1/2}$ with $M_n=M+e_Z-2n\beta|eB|/\hbar$, up to a common shift of order $\beta|eB|$, and $\langle\sigma_z\rangle=M_n/E_{n,\pm}$. The zeroth level has $\langle\sigma_z\rangle=-\sgn(eB)$, which equals $M/E_0$. The field therefore does not change the local spin signal, while the \emph{position} of the unpolarized zeroth level carries the sign.

\emph{Hall sign (Streda).} For $eB>0$ and $\mu$ in the gap, the zeroth level ($|eB|/h$ states per area) is filled if $M>0$ and empty if $M<0$. All other levels come in $\pm$ pairs. Relative to the charge-neutral reference,
\begin{equation}
n(\mu)=\frac{\sgn(M)}{2}\frac{eB}{h},\qquad \sigma_{xy}=e\frac{\partial n}{\partial B}\Big|_\mu=\frac{e^2}{2h}\sgn M,
\label{S-eq:streda}
\end{equation}
by Streda's formula~\cite{Streda1982}, consistent with the half-quantized surface Hall conductance~\cite{Qi2008,Nomura2011} and with $\sgn\Omega_{\rm occ}=\sgn M$.

\emph{Weak field.} The midpoint between $E_0$ and the nearest $n=1$ level is shifted by $\sgn(eBM)\,\hbar v^2|eB|/2|M|=\sgn(eBM)(\hbar v)^2/(2|M|\ell_B^2)$, with $\ell_B=\sqrt{\hbar/|eB|}=25.66$~nm$/\sqrt{B[\mathrm T]}$. For $v=4\times10^5$~m/s ($\hbar v=2.63$~eV\,\AA) and $|M|=15$~meV this is $3.5$~meV at 1~T ($2.0$~meV for $\hbar v=2$~eV\,\AA). The level spacing is $\sqrt2\hbar v/\ell_B=14.5$~meV\,$\sqrt{B[\mathrm T]}$ for $v=4\times10^5$~m/s. The Zeeman mass $g^*\mu_BB/2$ is $0.58$~meV at 1~T for $g^*=20$.

\emph{Magnetization shift of the band edges.} The midpoint shift is the orbital magnetic moment of the band edge. In weak fields each band edge moves as $E_n\to E_n-m_nB$, and in the two-band surface both edges carry the same moment, $|m|=e v^2/2|M|$ in units with $\hbar=1$, so the gap moves as a whole while its size is unchanged. The shift therefore survives when the Landau levels are not resolved. With a broadening of two and four times the level spacing near the edge, the field-odd shift of the half-height band edges is 0.90 and 0.81 of the formula above ($v=1$, $M=0.3$, $eB=0.004$). The Zeeman mass reverses with $B$ and changes the gap size. Its effect on the midpoint is of second order, $0.2\%$ for $g^*\mu_BB/2=0.05|M|$. A particle--hole-asymmetric term $Dk^2$ shifts all levels evenly in $B$ and drops out of the field-odd part. The mass curvature $\beta$ changes the shift by a relative amount of order $\beta|M|/v^2$, $-0.6\%$ and $-2.8\%$ for $\beta=0.02$ and $0.1$.

\emph{Orbital purity.} The equal moments of the two edges, and their link to the Berry curvature, are exact only for an isolated two-band edge. We tested the link in the multi-orbital $C_{3v}$ model of Sec.~\ref{S-sec:selection} with 4000 random parameter sets. For each set we computed the orbital moments of the two band-edge states at $\bar\Gamma$ from the full Bloch Hamiltonian, and added the atomic Zeeman term $\mu_B(L_z+2S_z)$ for a lattice constant of 4.14~\AA\ and an energy unit of 1~eV. The sign of the predicted field-odd midpoint shift equals the Hall sign in all 3993 parameter sets with a gapped edge, including sets whose edges mix $p_z$ and $p_\pm$ strongly. The moments of the two edges differ by $2.7\%$ (median). The two exceptions have a vanishing gap, below $10^{-5}$, where neither sign is defined. Larger deviations are possible when remote bands carry large orbital moments of their own, so the band edge should be checked for orbital purity in a given material.

\emph{Lattice check.} Model F1 on a strip in the Landau gauge (flux $\phi$ per plaquette, $\phi>0\Leftrightarrow eB>0$) has $E_0=-\sgn(\phi)(M-\pi t_2|\phi|)+O(\phi^2)$. Here $\beta=t_2/2$ and $|eB|/\hbar=2\pi|\phi|$ in lattice units. For $M=0.3$ and $\phi=0.01$ we find $E_0=-0.26882$ against $-0.26858$. By $\Theta_\perp$, $E_n(-\phi)=-E_n(\phi)$ holds exactly on the lattice, and a truncated-Fock diagonalization reproduces Eq.~\eqref{S-eq:E0} for both signs of $v$ to $10^{-16}$.

\begin{figure}[t]
\includegraphics[width=\textwidth]{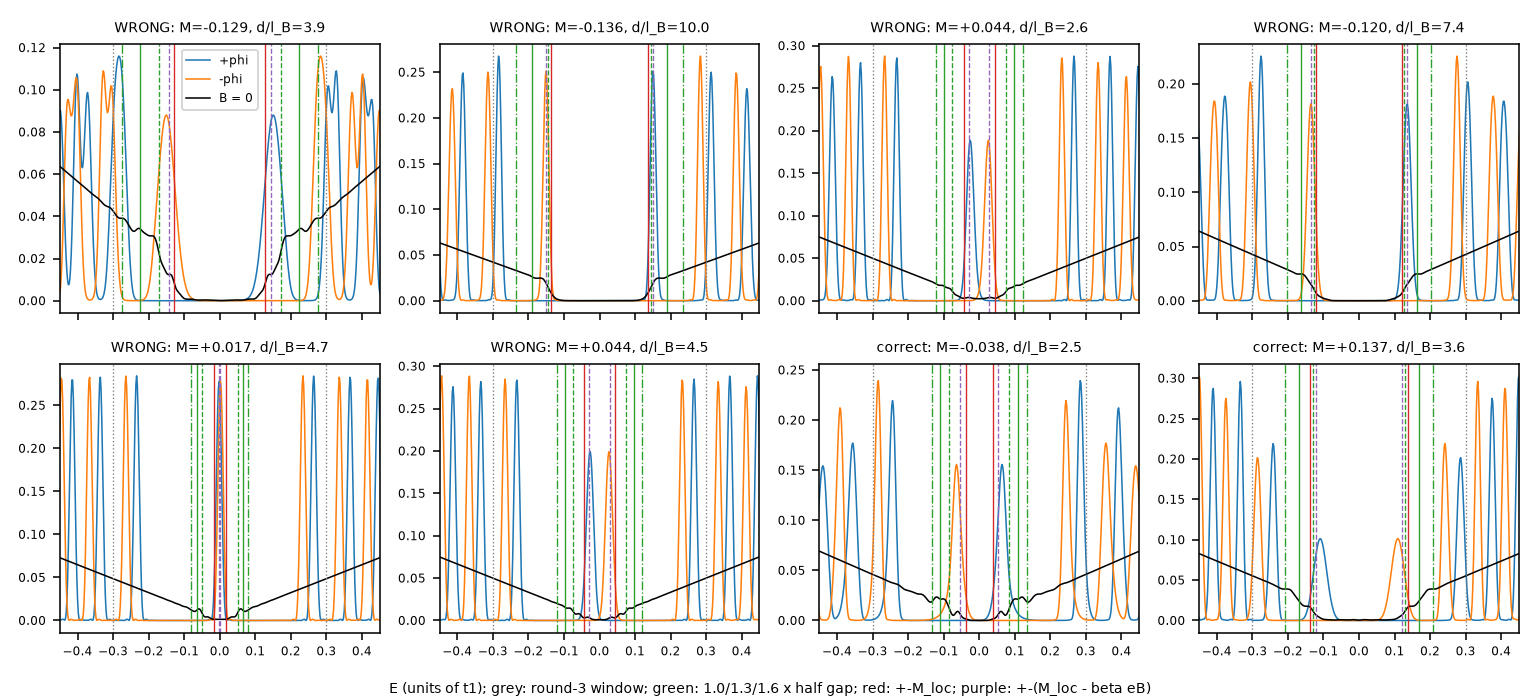}
\caption{Weak-field spectra on mixed-sign landscapes ($\ell_B=5.6$ sites). $\rho_0(E,+B)$ (blue), $\rho_0(E,-B)$ (orange) and $\rho_0(E,0)$ (black) at six far sites where the fixed-window sign assignment is wrong and two where it is correct. Grey dotted: the fixed window $|E|<0.3$, green: windows of 1.0, 1.3 and 1.6 zero-field half gaps, red: $\pm M_{\rm loc}$, purple: $\pm(M_{\rm loc}-\beta|eB|/\hbar)$.}
\label{S-fig:r4}
\end{figure}

\emph{Disorder.} With correlated mass disorder ($\xi=3$ and 10 sites, $\delta M_{\rm rms}\le0.3M_0$, $L=400$) and the zeroth level identified as the resolved $\rho_0$ peak nearest to $E=0$, the sign rule holds at every site for $\ell_B=4$ sites ($\ell_B/\xi=0.4$), at $\ge96.9\%$ of the sites for $\ell_B=5.6$, and at 75--97\% for $\ell_B=8$ sites, where disorder smears the zeroth level so that some sites show no resolved peak. Without disorder the estimator returns $E_0$ within $1.5\times10^{-3}$ of the lattice values. A single-field peak identification is therefore reliable only for $\ell_B\lesssim0.4\,\xi$. The field-odd sign assignment of the $\pm B$ protocol, $\sgn M=-\sgn\int_{|E|<1.3M_0}E\,[\rho_0(E,+B)-\rho_0(E,-B)]\,dE$, needs no resolved peak, because the levels $n\ge1$ are even in $B$ and cancel. It identifies the local mass sign at every site for $\ell_B=2.8$--8 sites and $\delta M_{\rm rms}\le0.3M_0$ (96 ensembles), including the 36 ensembles in which the $n=0\to1$ gap is smaller than twice the disorder, whereas the single-field peak picker falls to 47\% at $\ell_B=8$. On mixed-sign landscapes ($\delta M_{\rm rms}=0.1$, 31--50\% negative sites, $\xi=10$--40 sites, window $|E|<0.3$) the sign assignment is correct at 0.54--0.65 of the sites within $\ell_B$ of a sign boundary and at 0.87--0.98 of those farther than $2\ell_B$. The stored per-site data explain most of the deficit at strong fields. At $\ell_B=2.8$ and 4 sites the sign assignments agree with $\sgn(M-\beta|eB|/\hbar)$ at 0.993 and 0.988 of the far sites, and 39 of 40 (4 of 4) errors occur at sites with $0<M<\beta|eB|/\hbar$. There the curvature $\beta=t_2/2$ of the F1 regulator moves the zeroth level across zero, exactly as Eq.~\eqref{S-eq:E0} prescribes. On a physical surface $\beta$ is odd in the magnetization and small, so this regime does not arise; in the lattice it does because $\ell_B$ is only a few lattice constants. At $\ell_B=5.6$ sites the fraction was 0.875 and not explained by the curvature term. Further calculations on the identical landscapes and sites settled both points. Halving the regulator curvature ($t_2=1\to0.5$, $\beta=0.5\to0.25$) at $\ell_B=2.8$ sites leaves the sign assignments in exact agreement with $\sgn(M-\beta|eB|/\hbar)$ (1.000 of the 228 far sites at both curvatures) and raises the fraction correct against $\sgn M$ from 0.855 to 0.956, so the curvature term is the cause. At $\ell_B=5.6$ sites the stored spectra show that the fixed window $|E|<0.3$ admits several $n\ge1$ levels whose energies at $+B$ and $-B$ differ on these inhomogeneous landscapes; they do not cancel and can outweigh the zeroth level (Fig.~\ref{S-fig:r4}). Integrating $E\,[\rho_0(E,+B)-\rho_0(E,-B)]$ only over $|E|<w\,\Delta_{\rm loc}$, with $\Delta_{\rm loc}$ the zero-field half gap measured at the same site, gives the correct sign at every far site for $w=1$ and 1.3 (120 sites), and at 0.975 for $w=1.6$, against 0.842 with the fixed window. The field-odd rule therefore requires a window tied to the local gap.

\section{Domain walls}\label{S-sec:walls}
\emph{Jackiw--Rebbi mode.} For $H=v(-i\partial_x\sigma_y-k_y\sigma_x)+M(x)\sigma_z$ with $M(\pm\infty)=\pm M_0\,\sgn(\partial_xM)$, the $k_y=0$ zero mode obeys $\partial_x\psi=(M(x)/v)\sigma_x\psi$, where we multiplied by $\sigma_y$ and used $\sigma_y\sigma_z=i\sigma_x$. For $\sigma_x=s$ the solution is $\psi\propto e^{s\int^xM/v}$, which is normalizable only for $s=-\sgn(v)\sgn(\partial_xM)$. At finite $k_y$ the term $-vk_y\sigma_x$ gives
\begin{equation}
\sigma_x=-\sgn(v\,\partial_xM),\qquad E=|v|k_y\,\sgn(\partial_xM),
\label{S-eq:JR}
\end{equation}
so the mode's spin is perpendicular to the wall, it has $\langle\sigma_z\rangle=0$, and its group velocity is $|v|\sgn(\partial_xM)\hat y$. For a domain with $M>0$ (that is, $\Omega_{\rm occ}>0$) embedded in $M<0$, the outward normal $\hat x$ has $\partial_xM<0$, so the mode moves along $-\hat y=-\hat z\times\hat x$. It circulates clockwise when viewed from the vacuum side. This characterization is convention free.

\emph{Exact antisymmetry.} Uniform domains of opposite magnetization are exact time-reversal partners (the field-reversal relation of Sec.~\ref{S-sec:kramers}). For a wall with $M(-x)=-M(x)$, the mirror $M_x=i\sigma_x$ ($x\to-x$, $\sigma_x\to\sigma_x$, $\sigma_{y,z}\to-\sigma_{y,z}$) maps $v(k_x\sigma_y-k_y\sigma_x)$ to itself and $M(x)\sigma_z$ to $-M(-x)\sigma_z=M(x)\sigma_z$. Hence $\rho_z(x)=-\rho_z(-x)$, $\rho_0(x)=\rho_0(-x)$, and $\rho_x(x)=\rho_x(-x)$, within each $k_y$ sector.

\emph{Doubler-free continuum method.} A single Dirac cone cannot be regularized on a two-dimensional lattice without breaking time reversal, because the Wilson mass is necessarily $\propto\sigma_z$. We therefore solve Eq.~\eqref{S-eq:H} in a plane-wave basis along $x$ (period $L$, $N$ waves $q_n=2\pi n/L$), with matrix elements $\langle q_n|M|q_{n'}\rangle=\tilde M(q_n-q_{n'})$ computed by FFT of $M(x)$, and integrate $k_y\in[-k_c,k_c]$ numerically. The kinetic term is exact in this basis, so there are no doublers. The cutoff $vq_{\max}=4$ lies far above the energies shown. For \Mfig{walls} we use $L=300$, $N=384$, $k_c=2.5$, 201 values of $k_y$, $M_0=0.3$, $w=3$, and $\eta=0.03$. The exact antisymmetry holds to $3\times10^{-15}$, the reversed configuration gives $-\rho_z$ to $7\times10^{-15}$, and the Jackiw--Rebbi spin has the sign of Eq.~\eqref{S-eq:JR} at both walls. The in-gap density at the wall exceeds the bulk tail by a factor of $5.4$, and the 10--90\% width of the sign change is $2.25\,v/M_0$.

\emph{Lattice artefact.} For model F1 across the same walls, $A=+0.57$ inside and $-0.77$ outside at $E=1.2M_0$. Uniform F1 domains with $(m,t_2)$ and $(m',t_2)=(-M_0-2t_2,t_2)$ give $+0.637$ and $-0.803$ at $E=0.36$, but the exact time-reversal image $(-m,-t_2)$ gives $-0.637$. The asymmetry is therefore entirely due to the time-reversal-odd regulator. The ``band-inversion'' sign change of $A$ inside one domain has the same origin.

\begin{figure}[t]
\includegraphics[width=\textwidth]{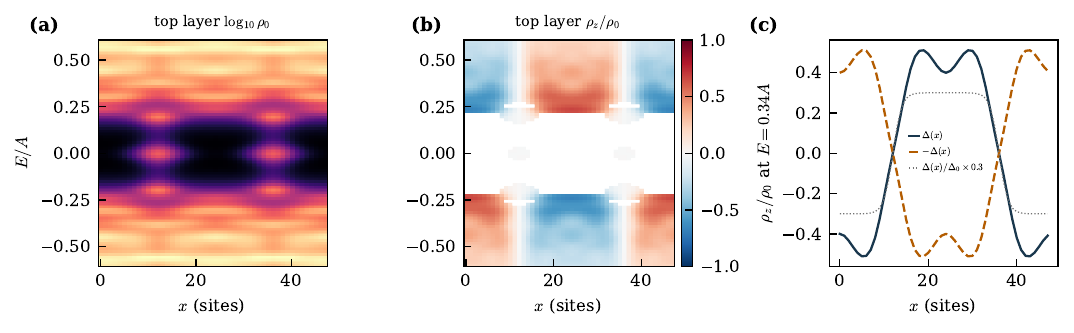}
\caption{Anomaly-free check with a topological-insulator slab ($N_z=6$) whose top layer carries two magnetic domain walls. (a) Top-layer $\log_{10}\rho_0$. The surface gap $\approx\pm0.22A$ and the in-gap wall modes. (b) Top-layer $\rho_z/\rho_0$. (c) Line cuts at $E=0.34A$ for $\Delta(x)$ and $-\Delta(x)$, which are exact negatives.}
\label{S-fig:slab}
\end{figure}

\emph{Anomaly-free slab.} The cubic four-band model $H=A\sum_i\sin k_i\Gamma_i+[M-2B\sum_i(1-\cos k_i)]\Gamma_4$, with $\Gamma_{1,2,3}=\tau_xs_{x,y,z}$ and $\Gamma_4=\tau_z$, is a strong topological insulator for $0<M<4B$. It is cut into $N_z$ layers, and an exchange $\Delta(x)s_z$ is applied to the top layer. The bulk is time-reversal invariant, so $\Delta(x)$ and $-\Delta(x)$ are exact partners without any two-dimensional regularization. For $A=B=M=1$, $N_x=48$, $N_z=6$, $\Delta_0=0.3$, and 32 values of $k_y$, the top-layer signals of the two configurations are exact negatives to $4\times10^{-14}$ (Fig.~\ref{S-fig:slab}). The domain values at $E=0.35A$ are $+0.40$ and $-0.43$. The small difference reflects the finite width of the strip and the $k_y$ sampling at the wall positions, not a breaking of the partner relation, which holds exactly.

\emph{Wall-mode spin in slabs.} Over 162 cubic slabs ($N_x\le96$, $N_z\le14$) the time-reversal residual is $\le1.7\times10^{-12}$, the domain values are antisymmetric within $0.3\%$, and the in-gap density at the walls is enhanced by 1.4--13. In the cubic model the $z$-surface states are eigenstates of $\tau_ys_z$, which anticommutes with $\tau_0s_{x,y}$. The physical in-plane spin therefore has no surface projection, and the wall-mode spin is carried by $\tau_xs_y$. A rerun of all 162 slabs with this operator gives opposite spins at the two walls in every slab, with magnitude $\ge0.62$ for $N_z\ge10$ (up to 0.94), directed along the wall because the surface texture of this model is radial. In a Bi$_2$Se$_3$-type model with the helical in-plane term $A(\sin k_y\tau_xs_x-\sin k_x\tau_xs_y)$ and a spinless $k_z$ term $\tau_y$, whose surface states are eigenstates of $\tau_x$, the physical spin $\tau_0s_x$, perpendicular to the wall, takes opposite values at the two walls in all 16 slabs ($N_x=48$--72, $N_z=10$--14, $\Delta_0=0.2$--0.3), with magnitude 0.85--0.91, while $\tau_xs_y$ vanishes there, the reverse of the cubic model. The time-reversal partners agree to $5\times10^{-13}$, and the in-gap enhancement is 3.7--9.2. On random mixed-sign landscapes the boundary modes carry the same spin, aligned with $-\nabla M$ (Sec.~\ref{S-sec:numerics}).

\section{Tip-induced mass}\label{S-sec:tip}
A magnetized tip perturbs the surface under the apex through direct exchange, the Zeeman term of its stray field and the orbital (vector-potential) term of the same field. All three reverse with the tip polarization. The exchange and Zeeman parts shift the local mass by $\delta M_{\rm tip}=cP$, and to first order the tip-reversal asymmetry becomes
\begin{equation}
A(E)=\frac{P\rho_z}{\rho_0}+\delta M_{\rm tip}\,h(E),\qquad h(E)=\frac{\partial_M\rho_0}{\rho_0},\qquad
\partial_M\rho_0=-\sgn(M)\frac{|M|}{2\pi v^2}\bigl[L_\eta(E-|M|)+L_\eta(E+|M|)\bigr],
\label{S-eq:tipmass}
\end{equation}
where the last expression follows from Eq.~\eqref{S-eq:rho0}. Because $\rho_0$ is even in $M$ (the field-reversal relation of Sec.~\ref{S-sec:kramers}), $\partial_M\rho_0$ is odd, so the artefact changes sign under domain reversal and under tip reversal, exactly like the signal, and it is unaffected by $\pm B$ averaging. Two symmetry arguments decide what can and cannot remove it.

\thm{Generalized Kramers null result.} If $\Theta_TH\Theta_T^{-1}=H$ and the tip perturbation satisfies $\Theta_TV_P\Theta_T^{-1}=V_{-P}$, then $G_{+P}(\mathbf r,E)=G_{-P}(\mathbf r,E)$ for all $\mathbf r$ and $E$.
$H+V_{-P}=\Theta_T(H+V_P)\Theta_T^{-1}$, and the covariance relation of Sec.~\ref{S-sec:sign} with $W=1-\mathbf P\cdot\bm\sigma$, for which $\Theta_T^{-1}W\Theta_T=1+\mathbf P\cdot\bm\sigma$, gives $G_{-P}=G_{+P}$.

All three channels satisfy the assumption. For the orbital channel this is because $\Theta_T$ conjugates the Peierls phases. On a nonmagnetic reference the tip-reversal asymmetry therefore vanishes even in the presence of the tip's own perturbation. The null test of the protocol checks the instrument, and the tip can be calibrated only through observables that are even in $P$.

\thm{Particle--hole parity.} Let $\Xi$ be antiunitary with $\Xi H\Xi^{-1}=-H$, $\Xi\sigma_z\Xi^{-1}=-\sigma_z$ and $\Xi\,1\,\Xi^{-1}=1$. For model F1 with any real mass map, and for Eq.~\eqref{S-eq:H} with a position-dependent mass, $\Xi=\sigma_xK$. If $V_P=\delta_P(\mathbf r)\sigma_z$ with any profile, then $\rho_0$ is even and $\rho_z$ odd in $E$ for both tip configurations, and the odd part $A_{\rm odd}(E)=[A(E)-A(-E)]/2$ differs from its unperturbed value only at order $\kappa\,\delta^2$.
$\Tr[\sigma_z\delta(E-H)]=\Tr[\Xi\sigma_z\Xi^{-1}\delta(E+H)]=-\Tr[\sigma_z\delta(E+H)]$, and similarly $\rho_0(E)=\rho_0(-E)$. With $\Delta$ and $\Sigma$ the difference and sum over the two tip configurations, $A=(\Delta\rho_0+\kappa\Sigma\rho_z)/(\Sigma\rho_0+\kappa\Delta\rho_z)$, where $\Delta\rho_0,\Sigma\rho_0$ are even and $\Sigma\rho_z,\Delta\rho_z$ are odd, so $A_{\rm odd}=\kappa[\Sigma\rho_z/\Sigma\rho_0-\Delta\rho_0\Delta\rho_z/(\Sigma\rho_0)^2]+O(\kappa^3)$ with $\Delta\rho_0,\Delta\rho_z=O(\delta)$.

The two-sided band-edge sign assignment, the difference of the windows above the upper and below the lower edge, uses only $A_{\rm odd}$. It is therefore protected at first order against exchange and Zeeman tip masses of any spatial profile. One-sided estimators, with all energies on one side of the gap, are not. Ignoring the artefact biases the one-sided $1/E$ estimate by $\delta M_{\rm tip}\sum_jh_j/E_j/S_2$, which for $\kappa=0.05$ is $-27\%$ at $\delta M_{\rm tip}=0.2|M|$ and reverses the sign at $0.73|M|$. Its lineshape is localized at the band edge, so a joint one-sided fit with the regressors $1/E$ and $h(E)$ removes the bias at a standard-error inflation $1/\sqrt{1-r^2}=1.35$ (\Mfig{stats}c), provided the lineshape is right. The protection is broken by scalar terms (a $Dk^2$ term, potentials), by in-plane Zeeman components, and by the orbital term, whose first-order contribution is odd because $\Xi$ reverses the field.

\emph{Lattice tests (model F1 on $256\times256$ sites with four boundary twists, 3072 moments, and $\kappa=0.2$).}
\begin{itemize}
\item \textbf{Parity.} For exchange footprints from one site to uniform and $J_0\le0.1M_0$, the odd part in the window $|E|\in[1.5,2.5]|M_0|$ changes by at most $0.33\%$, and for the Zeeman part of a dipolar stray field by $0.16\%$. A particle--hole-breaking $Dk^2$ term of about 8~eV\,\AA$^2$ raises this to $1.9\%$. The sign survives up to $J_0=M_0$ (odd ratio $\ge0.49$) and is lost only at $J_0=3M_0$ with large footprints.
\item \textbf{Orbital stray field.} The deviation of the odd part is linear in the field below the apex. The ratio between 0.5 and 0.25~T is 2.00--2.01. At $\kappa=0.2$ the near-edge sign assignment reverses above $B^*=1.75$, 1.36 and 1.45~T for apex heights of 13, 27 and 53~nm, and not below 2~T for 6.7~nm. Because the orbital term does not scale with $\kappa$ while the signal does, $B^*\propto\kappa$. About 0.4~T at $\kappa=0.05$.
\item \textbf{Switching.} In Voronoi patch landscapes with log-normal coercive fields, measured with the sequence $+P,-P,+P,-P$, the tip field switched the patch under the tip at 22 of 576 sites. The asymmetry did not vanish. It was reproducible in both pairs, and 12 of the 20 decided switched sites gave the wrong sign, because neighbouring patches do not follow the tip. All 223 unswitched sites farther than 10 sites from a patch boundary were correct. Switching must therefore be excluded rather than detected. Keep the tip field below the local coercive field, and compare gap maps taken with a nonmagnetic tip before and after.
\end{itemize}
An antiferromagnetic tip has no stray field and removes both the orbital term and switching. The remaining exchange mass then cancels in the two-sided sign assignment.

\section{Statistics of the sign assignment}\label{S-sec:stats}
For $A_j=aE_j^{-1}+\epsilon_j$ with independent Gaussian noise of variance $\sigma^2$, the best linear unbiased estimator is $\hat a=S_2^{-1}\sum_jA_j/E_j$ with standard error $\sigma/\sqrt{S_2}$, where $S_2=\sum_jE_j^{-2}$. A two-sided test at level $\alpha$ reports $\sgn\hat a$ when $|\hat a|\sqrt{S_2}/\sigma>z_{1-\alpha/2}$. The probability of the correct report is $\Phi(\mu-z_{1-\alpha/2})$ with $\mu=|a|\sqrt{S_2}/\sigma$ and $a=\kappa M$, and the probability of a wrong-sign report is $\Phi(-\mu-z_{1-\alpha/2})$, which is negligible. To reach $P_{\rm correct}\ge0.95$ at $\alpha=0.01$ requires $|\kappa M|\sqrt{S_2}/\sigma\ge z_{0.995}+z_{0.95}=4.22$. For $M=15$~meV and 60 energies in $[1.3,4]|M|$, $\sqrt{S_2}=0.226$~meV$^{-1}$, so with $\kappa=0.05$ the requirement is $\sigma\le0.040\,G_0$. The closed form ($0.193$ at $\sigma=0.1G_0$) agrees with Monte Carlo ($0.19$). The $\pm B$ protocol averages two independent spectra per energy, which halves the variance, and it cancels the Zeeman mass exactly. For $M=0.4$~meV and an opposing Zeeman mass of $0.58$~meV, all sign assignments are wrong without averaging and all are right with it. In the boost regime of Sec.~\ref{S-sec:selection}, the regressor set is extended to $\{1/E,\,1-M^2/E^2,\,h(E)\}$ or the window is restricted to $|E|\lesssim|M|/|\delta|$.

The biases quoted in Sec.~\ref{S-sec:tip} and in the main text refer to one-sided estimators, with all energies on one side of the gap. With energies on both sides, five estimators can be compared:
\begin{itemize}
\item the one-sided $1/E$ fit
\item the one-sided joint fit with $\{1/E,h(E)\}$
\item the two-sided fit to the odd signal lineshape
\item a two-sided joint fit that adds an odd, edge-localized regressor for the orbital term
\item a height-joint fit across five tip heights with per-height artefact amplitudes.
\end{itemize}
Table~\ref{S-tab:estimators} gives their wrong-sign rates in synthetic data. The data use closed-form lineshapes, 30 energies per side in $[1.05,4]|M|$ and noise 0.02 per point, with mass-type, particle--hole-leak and orbital artefacts, and sign assignments at $3\sigma$ (256 settings, 2000 draws each). The two-sided estimator is protected against mass-type artefacts even when their lineshape is twice as broad as assumed, but fails when a strong orbital term meets a weak tip polarization. The two-sided joint and height-joint estimators, whose regressors match the model's lineshapes, are never wrong.

\begin{table}[h]
\caption{Worst wrong-sign rate over groups of synthetic settings (mean undecided fraction in brackets).}
\label{S-tab:estimators}
\begin{ruledtabular}
\begin{tabular}{lcccccc}
Group & settings & one-sided & one-sided joint & two-sided & two-sided joint & height joint\\ \hline
no orbital term & 64 & 1.000 (0.14) & 1.000 (0.08) & 0.003 (0.06) & 0.000 (0.00) & 0.000 (0.00)\\
orbital $\ge0.1$, $\kappa=0.05$ & 64 & 1.000 (0.14) & 1.000 (0.18) & 1.000 (0.21) & 0.000 (0.06) & 0.000 (0.01)\\
orbital $\ge0.1$, $\kappa=0.2$ & 64 & 1.000 (0.06) & 0.000 (0.01) & 0.000 (0.00) & 0.000 (0.00) & 0.000 (0.00)\\
$\delta_0=|M|$, no orbital & 16 & 1.000 (0.00) & 1.000 (0.00) & 0.003 (0.24) & 0.000 (0.00) & 0.000 (0.00)\\
\end{tabular}
\end{ruledtabular}
\end{table}

\section{Quasiparticle interference with the exact $T$ matrix}\label{S-sec:qpi}
For model F1 on an $N\times N$ torus, $G(\mathbf k;z)=(z-d_0+\mathbf d\cdot\bsig)/[(z-d_0)^2-|\mathbf d|^2]$ and $G(\mathbf R)=N^{-2}\sum_{\mathbf k}e^{i\mathbf k\cdot\mathbf R}G(\mathbf k)$. A defect $U$ at the origin has $T=U(\id-g_0U)^{-1}$ with $g_0=G(0)$, and $\delta G(\mathbf R,\mathbf R)=G(\mathbf R)TG(-\mathbf R)$. The map in channel $W$ is $\delta\rho_W=-\pi^{-1}\mathrm{Im}\Tr[W\delta G]$. By the covariance relation of Sec.~\ref{S-sec:sign} the maps of $(t_1,m,t_2)$ and of its images under Table~\ref{S-tab:action}, with images of $W$ and $U$, coincide. For a generic defect $U=(0.8,0.3,-0.2,0.5)$ and four tips the differences are $2\times10^{-17}$ ($\Theta_\perp$), $3\times10^{-17}$ ($\Theta_T$), and $0$ ($R_z$), against a signal of $7\times10^{-2}$. The Born approximation becomes inaccurate at moderate defect strength, so quantitative inversions use the exact $T$ matrix.

\section{Numerical methods and tests}\label{S-sec:numerics}
All calculations use our Python package \texttt{spdirac}, which implements the continuum and lattice models, exact $T$-matrix maps, exact diagonalization, kernel-polynomial (KPM) routines~\cite{Weisse2006}, and the statistics. Each analytic result was tested against an independent numerical calculation, 22 tests in total, and all agree. The record of these tests is kept with the code.

\begin{figure}[t]
\includegraphics[width=\textwidth]{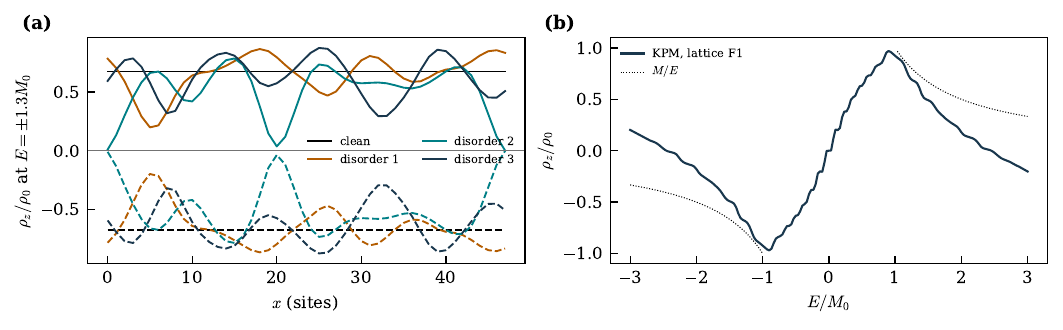}
\caption{KPM test. (a) $\rho_z/\rho_0$ at $E=+1.3M_0$ (solid) and $-1.3M_0$ (dashed) along a line cut of a $48\times48$ F1 lattice, clean and for three mass-disorder realizations. (b) Clean local spectrum compared with $M/E$. The deviation at large $|E|$ is the regulator effect of Sec.~\ref{S-sec:walls}.}
\label{S-fig:kpm}
\end{figure}

\emph{KPM test.} The KPM moments of the local $2\times2$ spin block agree with exact spectral moments to $3\times10^{-15}$. On a $48\times48$ F1 lattice with Gaussian-correlated mass disorder (correlation length 3 sites, $\delta M_{\rm rms}=0.3M_0$) and 700 moments, the local sign of $\rho_z/\rho_0$ at $E=\pm1.3M_0$ is preserved at every site of a line cut in three realizations. The amplitude, however, fluctuates by up to $100\%$, with a minimum of $0.001$ (Fig.~\ref{S-fig:kpm}). Ensemble statistics are therefore required.

\emph{Disorder ensembles.} On F1 lattices up to $1024\times1024$ ($M_0=0.3$, 64 sites per realization, 1024 moments) the fraction of sites with the correct sign of $\rho_z/\rho_0$ at the fixed energies $\pm1.2$, $\pm1.5$ and $\pm2.0\,M_0$ is 1.00 for $\delta M_{\rm rms}=0.1M_0$ with $\xi\ge5$ sites (0.98 at $\xi=2$), 0.88--0.95 at $0.2M_0$, and 0.70--0.86 at $0.3M_0$. It does not depend on $L$ and decreases away from the band edge (minima 0.94, 0.92 and 0.70 at the three energies). Charge puddles of $0.3M_0$ lower it to 0.61--0.89. Negative local masses occur at $\le0.6\%$ of the sites, so most wrong signs are not sign changes of $M$. Two effects contribute. A fixed energy lies inside the local gap wherever the local mass exceeds it, and the F1 regulator reverses $\langle\sigma_z\rangle$ near $E^*\simeq t_1\sqrt{2M_0/t_2}\approx2.6M_0$ (Sec.~\ref{S-sec:walls}), a threshold that local reductions of $M$ lower. Sign maps should therefore be taken at each site's own band edge, as in the protocol of the main text.

\emph{Reading the sign at each site's band edge.} For each site the local band edges are located at the steepest rise of $\rho_0$, and the sign is decided from the mean $\rho_z/\rho_0$ in a window of $0.3M_0$ beyond both edges. On $384\times384$ F1 lattices (96 sites per ensemble, 2048 moments) this sign assignment is correct at every site for $\delta M_{\rm rms}\le0.3M_0$ and $\xi=3$--10 sites, with or without puddles of $0.3M_0$, at $M_0=0.1$, where the regulator's reversal lies at $E^*\approx4.5M_0$ (60 ensembles). At $M_0=0.3$ it is correct at $\ge99.0\%$ of the sites (60 ensembles), while fixed-energy sign assignments on the same spectra fall to 0.74--0.94. These ensembles contain no sites with negative local mass, but the sign assignment cannot be biased toward one sign. Under the exact time-reversal image $(m,t_2)\to(-m,-t_2)$ every spectrum maps to its reverse (the covariance relation of Sec.~\ref{S-sec:sign}) and each sign assignment flips. A control with the mass field reversed at fixed regulator, which is the trivial phase of F1, gives 0.91--1.00 at $\delta M_{\rm rms}=0.3M_0$. 

\emph{Mixed-sign landscapes.} For $M(\mathbf r)=\bar M+\delta M\,g(\mathbf r)$ with $\delta M=0.1$, $\bar M=0$ and 0.05 (50\% and about 31\% negative sites) and $\xi=5$--40 sites, the same band-edge sign assignment is correct at 0.998 of sites with positive and 1.000 of sites with negative local mass farther than 10 sites (about one Dirac length $v/\delta M$) from the nearest sign boundary, for $\xi\ge20$ (1026 sites). The zone in which the fraction correct is below 0.95 is 5--10 sites wide (0.5--1 Dirac lengths). Within 1--2 sites of a boundary the fraction correct is 0.55--0.61, and only 23--74\% of the sites give a confident sign assignment. The boundaries are marked instead by their Jackiw--Rebbi modes: at $E=0$ and $d\le1.5$ sites the median in-plane polarization $|\rho_\parallel|/\rho_0$ is 0.81--0.99, and its median alignment with $-\nabla M$ is 0.86--0.99 (0.97--0.99 for $\xi\ge20$). In 16 exact time-reversal images every sign assignment flips and $\rho_0$ agrees to $3.5\times10^{-15}$. Sign-reversed patches larger than about two Dirac lengths therefore contain an interior in which the sign is imaged reliably, and an in-plane tip outlines their boundaries.

\section{Large-scale simulations}\label{S-sec:campaigns}
We tested the analytic results in about 3800 large lattice calculations on a computing cluster. For each test the criterion for agreement was fixed before the calculation. Secs.~\ref{S-sec:sign}--\ref{S-sec:numerics} give the details, and a record of all tests and their outcomes is kept with the code.

Most tests confirmed the analytic results directly. Three first failed for reasons unrelated to the physics and were repeated. A partner map must image a defect moment of unknown sign (Sec.~\ref{S-sec:sign}). The physical spin $\tau_0s_x$ has no surface projection in the cubic slab (Sec.~\ref{S-sec:walls}). A fixed energy window for the zeroth Landau level admits higher levels at weak field (Sec.~\ref{S-sec:landau}). Two outcomes changed the protocol. Sign maps taken at a fixed energy degrade under disorder, so the sign must be read at each site's own band edge (Sec.~\ref{S-sec:numerics}). A tip field strong enough to switch the sample can give a reproducible wrong sign, so switching must be excluded (Sec.~\ref{S-sec:tip}).

\end{document}